\documentclass{emulateapj}
\usepackage{graphicx}
\usepackage{epstopdf}
\usepackage{epsfig}

\newcommand{\hmpc}{\ensuremath{\,h^{-1}\,{\rm Mpc}\,}}

\newcommand{\avg}[1]{\ensuremath{\langle #1 \rangle}}
\newcommand{\bma}{\begin{math}}
\newcommand{\ema}{\end{math}}
\newcommand{\beq}{\begin{equation}}
\newcommand{\eeq}{\end{equation}}
\newcommand{\beqa}{\begin{eqnarray}}
\newcommand{\eeqa}{\end{eqnarray}}
\newcommand{\bc}{\begin{center}}
\newcommand{\ec}{\end{center}} 
\newcommand{\bit}{\begin{itemize}}
\newcommand{\eit}{\end{itemize}}

\font\BFd=cmmib10
\font\BFt=cmmib10
\font\BFs=cmmib10 scaled 700
\font\BFss=cmmib10 scaled 500

\def\bbox#1{%
\relax\ifmmode
\mathchoice
{{\hbox{\BFd #1}}}
{{\hbox{\BFt #1}}}
{{\hbox{\BFs #1}}}
{{\hbox{\BFss #1}}}
\else \mbox{#1} \fi }

\def\r{{\bbox{r}}}

\begin{document}



 
\submitted{\today. To be submitted to \apj.} 

\title{Intensity Mapping with Carbon Monoxide Emission Lines and the Redshifted 21 cm Line}
\author{Adam Lidz\altaffilmark{1}, Steven R. Furlanetto\altaffilmark{2}, S. Peng Oh\altaffilmark{3}, James Aguirre\altaffilmark{1}, Tzu-Ching Chang\altaffilmark{4}, Olivier Dor\'e\altaffilmark{5,6},
Jonathan R. Pritchard\altaffilmark{7}}
\altaffiltext{1} {Department of Physics \& Astronomy, University of Pennsylvania, 209 South 33rd Street, Philadelphia, PA 19104, USA}
\altaffiltext{2}{Department of Physics \& Astronomy, University of California Los Angeles; Los Angeles, CA 90095, USA}
\altaffiltext{3}{Department of Physics, University of California, Santa Barbara, CA 93106, USA}
\altaffiltext{4}{IAA, Academia Sinica, P.O. Box 23-141, Taipei 10617, Taiwan}
\altaffiltext{5}{Jet Propulsion Laboratory, California Institute for Technology, Pasadena, CA 91109}
\altaffiltext{6}{California Institute of Technology, Pasadena, CA 91125}
\altaffiltext{7}{Harvard-Smithsonian Center for Astrophysics, Cambridge, MA 02138, USA}
\email{alidz@sas.upenn.edu}

\begin{abstract}
We quantify the prospects for using emission lines from rotational transitions of the CO molecule to perform an `intensity mapping' observation at high 
redshift during the Epoch of Reionization (EoR).  The aim of CO intensity mapping is to observe the combined CO emission from many unresolved galaxies,
to measure the spatial fluctuations in this emission, and use this as a tracer of large scale structure at very early times in the history of our Universe. This measurement
would help determine the properties of molecular clouds -- the sites of star formation -- in the very galaxies that reionize the Universe.
We further
consider the possibility of cross-correlating CO intensity maps with future observations of the redshifted 21 cm line. The cross spectrum is less sensitive to foreground
contamination than the auto power spectra, and can therefore help confirm the high redshift origin of each signal. 
Furthermore, the cross spectrum measurement would help extract key information about the EoR, especially regarding the size distribution of ionized regions.
We discuss uncertainties in predicting the CO signal at high redshift, and discuss strategies for improving these predictions. Under favorable assumptions, and feasible
specifications for a CO survey mapping the CO(2-1) and CO(1-0) lines, the power spectrum
of CO emission fluctuations and its cross spectrum with future 21 cm measurements from the MWA are detectable at high significance.
\end{abstract}


\section{Introduction} \label{sec:intro}

In this paper, we consider the possibility of studying large scale structure at very high redshift ($z \gtrsim 6$) using rotational emission lines from 
CO molecules, as first proposed by Righi et al. (2008). We further consider cross-correlating these measurements with upcoming data from
redshifted 21 cm line surveys. In this introductory section, we first motivate our study by discussing the scientific promise of these measurements.

One of the primary goals of observational cosmology at present is to detect, and elucidate the nature of, the Epoch of Reionization (EoR). The EoR is a key
stage in the history of our Universe when early galaxies and quasars turn on and photo-ionize `bubbles' of neutral hydrogen gas in their surroundings. The ionized bubbles
grow and merge, eventually filling essentially the entire volume of the intergalactic medium (IGM) with ionized gas.  Detailed measurements of the EoR will help determine the properties of
the first galaxies and the nature of the high redshift IGM, thereby providing important observational input for theories of first structure formation. 

The most direct, and perhaps the most promising, way of studying the EoR is to use the redshifted 21 cm line, which may ultimately provide full three-dimensional maps of the reionization
process (e.g. Madau et al. 1997, Zaldarriaga et al. 2004, Furlanetto et al. 2006a).  Motivated by the promise of this technique, a first generation of experiments aimed at detecting the 21 cm signal from the EoR is
currently underway, including the Murchison Widefield Array (MWA) (Lonsdale et al. 2009); the Low Frequency Array (LOFAR) (Harker et al. 2010); Precision Array for Probing the Epoch of Reionization (PAPER) (Parsons et al. 2010); and the Giant Metrewave Radio Telescope (GMRT) (Paciga et al. 2010). These first experiments will detect the 21 cm signal only {\em statistically}, as they will not have the sensitivity to make detailed maps of the
EoR (McQuinn et al. 2006).  The surveys will measure, for example, the power spectrum of 21 cm fluctuations by binning together many individually noisy wave modes.
The experiments must overcome several observational challenges, especially systematic effects from strong galactic and extra-galactic foreground emission, coupled
with instrumental effects from the beam, polarization response, and calibration 
errors (e.g. Liu et al. 2009, Datta et al. 2010, Harker et al. 2010, Petrovic \& Oh 2010).  

Given the statistical nature of the observations, and the challenge of the measurements, it is natural to ask two related questions. First, how can one ensure that the measured signal
truly originates from the high redshift IGM and is not the result of, for example, residual foreground contamination? Second, how does one best interpret the statistical measurement and robustly infer key information about reionization? One way of potentially addressing both of these questions is to cross-correlate the redshifted 21 cm signal with a galaxy
survey at very high redshift (Furlanetto \& Lidz 2007, Wyithe \& Loeb 2007a, Lidz et al. 2009). This measurement is much less sensitive to foreground contamination than measurements of the 21 cm auto power spectrum since most of the anticipated 21 cm foregrounds come from low redshift -- primarily galactic synchrotron -- and therefore {\em do not} impact the high redshift galaxy survey. In addition, the cross spectrum provides a more direct tracer of bubble growth during reionization than the auto spectrum: the cross spectrum turns over on
scales smaller than the size of ionized 
bubbles around the galaxies in the survey (Lidz et al. 2009). Hence the cross spectrum can both confirm the high redshift origin of a putative redshifted 21 cm
signal, and facilitate its interpretation.  

The difficulty with this proposal is that high redshift galaxies are, of course, very difficult to detect. Further, in many respects the
galaxy and 21 cm surveys are ill-matched (Furlanetto \& Lidz 2007): galaxy surveys typically have high angular resolution but -- without expensive spectroscopic follow-up -- they have poor spectral resolution. On the other hand, the 21 cm surveys have high spectral resolution, yet poor angular resolution. Further, the purely transverse modes that are easiest to measure
with a galaxy survey are lost in the 21 cm data owing to foreground contamination/cleaning. Finally, galaxy surveys become increasingly challenging at progressively higher redshifts. 
Nevertheless, it may be possible to detect the cross spectrum using upcoming surveys for Lyman-alpha emitting galaxies (LAEs) with the Hyper-Suprime Cam on the Subaru
telescope (Lidz et al. 2009).
 
In this paper, we consider the complementary possibility of cross-correlating `intensity maps' of CO emission from high redshift galaxies 
with upcoming 21 cm data. Alongside studies of the 21 cm signal
from the EoR, it was pointed out that similar experiments could map large scale structure {\em after reionization} by detecting 21 cm emission from residual neutral hydrogen locked up
in galaxies and damped Ly-$\alpha$ systems (Chang et al. 2008, Wyithe \& Loeb 2007b). These works advocate an approach dubbed `intensity mapping', that is rather orthogonal to
that of traditional galaxy surveys. In the intensity mapping approach, one simply measures the large scale variations in 21 cm emission from many individually unresolved galaxies across
the Universe. This lets one measure large scale modes of interest (such as the baryon acoustic oscillation scale), without actually resolving individual galaxies, which in turn allows for an inexpensive experiment. 
One might wonder if intensity mapping observations may be usefully performed with {\em other (than 21 cm) emission lines}, and also at high redshifts during the EoR. 
Indeed, one such possibility was 
considered by Righi et al. (2008), who suggested performing an intensity mapping experiment using rotational emission lines from the CO molecules residing in star-forming 
galaxies at high redshift.\footnote{A few earlier papers suggested performing high redshift 
intensity mapping observations with various atomic fine
structure lines (Suginohara et al. 1999, Basu et al. 2004, Hern\'andez-Monteagudo et al. 2007, 2008).}  This possibility is in part motivated by detections of CO emission lines from
individual quasar host galaxies at $z \gtrsim 6$ (Walter et al. 2003).
A further intriguing possibility is then to cross-correlate a high redshift CO intensity map with upcoming 21 cm observations. In particular, note that
CO intensity mapping experiments may be better matched to future 21 cm observations than more traditional high redshift galaxy surveys.

Moreover, a CO intensity mapping
experiment also provides a unique large-scale view of star formation during the `cosmic dawn.' Observations from the Hubble Space
Telescope and ground-based telescopes are providing first glimpses into this era, and in the future new instruments like ALMA, JWST,
and large near-infrared ground-based telescopes will provide even more detail.  However, these instruments (and especially ALMA) are
all restricted by relatively small fields of view, requiring targeted surveys over small scales.  Intensity mapping complements such
efforts beautifully, providing maps on several-degree scales.  It will therefore provide an unbiased view of the distribution of
CO-emitting gas -- and hence a window into the distribution of star formation over a large range of environments -- that is difficult
to assemble from targeted measurements of individual galaxies. Together with the detail provided by ALMA, such a measurement will
provide a complete census of molecular emission during the EoR.

Here we aim to extend the work of Righi et al. (2008) and quantify the survey requirements for measuring the power spectrum of CO emission fluctuations, as
well as the 21 cm-CO cross power spectrum. One difference with this previous work is that these authors considered the angular power spectrum of CO fluctuations
as a function of spectral resolution. In this work, we instead treat the CO data in a fully three-dimensional way  --  translating observed frequency for gas emitting in a given CO line to 
line of sight distance (modulo peculiar velocities) -- as planned for future 21 cm data sets. As with the 21 cm data, one can clean foreground emission from the CO data 
by removing the spectrally smooth component of the emission in each spatial pixel. In this paper, we focus on two particular CO emission lines: the CO(1-0) line and the CO(2-1) line.
These two lines are convenient because their redshifted emission from the EoR can potentially be observed from the ground, i.e., the proposed observations do not
require a costly space-based mission.  For reference, gas at $z=7$ emitting in the CO(1-0) and
CO(2-1) lines can
be observed at $\nu = 14$ Ghz and $\nu = 29$ Ghz respectively. Having two CO emission line tracers of high redshift structure further guards against foreground contamination in
the CO data.   

The outline of this paper is as follows. In \S \ref{sec:co_bright}, we describe our model for star formation and the specific intensity of CO emission from high redshift
galaxies. \S \ref{sec:mean_co} calculates the spatially averaged CO emission from star-forming galaxies as a function of redshift. In \S
\ref{sec:co_flucs} we then use numerical simulations of high redshift structure formation and reionization (McQuinn et al. 2007a, 2007b) to compute the power spectrum of spatial fluctuations in the CO emission,
while \S \ref{sec:cross_co} computes the cross spectrum between the CO and 21 cm signals. \S \ref{sec:trans_co} discusses the impact of foreground emission on the CO measurements,
and describes the benefit of detecting emission from two different rotational lines from emitting gas at a given redshift. \S \ref{sec:detectability} then calculates the detectability of the auto
and cross spectrum signals for various plausible survey configurations. 
In \S \ref{sec:previous} we compare our results with other related calculations, especially closely related work from Carilli (2011) and Gong et al.
(2011). In addition to these recent papers and Righi et al. (2008), our work has some overlap with Visbal \& Loeb (2010), who recently considered cross-correlating various spectral lines (including CO lines) along
with the redshifted 21 cm line. Their work focused on studying large scale structure with these lines after reionization, and so our work which considers the EoR, is nicely complementary. Finally, we conclude in \S \ref{sec:conclusions} and mention possible follow-up work.
Throughout we consider a $\Lambda$CDM cosmology parametrized by $n_s =1, \sigma_8 = 0.8, \Omega_m = 0.27, \Omega_\Lambda = 0.73, \Omega_b = 0.046$, and $h=0.7$, (all symbols have their usual meanings), consistent with the latest WMAP constraints from Komatsu et al. (2010).

\section{Modeling the CO Brightness Temperature}
\label{sec:co_bright}

In order to model the CO emission from high redshift galaxies, we adopt a simple model to
connect the strength of this emission to the abundance of the dark mater halos that host
CO luminous galaxies. In brief, we first relate a galaxy's star formation rate (SFR) to the dark matter
mass of the galaxy's host halo. We then assume that empirical correlations between a galaxy's SFR and its luminosity
in the CO lines of interest, measured from galaxies at $z \lesssim 3$, apply also at higher redshifts during the EoR. 
Throughout, we consider how uncertainties in this model impact our theoretical predictions. 

First, let us consider the specific intensity of the CO emission.  If the proper volume emissivity of CO emitting galaxies
at redshift $z^\prime$, emitting at an observed frequency of $\nu_{\rm obs}$, 
is $\epsilon \left[\nu_{\rm obs}(1+z^\prime)\right]$, then the specific intensity of the observed CO emission can be
determined by solving the (absorption free) cosmological radiative transfer equation.
The solution at $z=0$ and observed frequency $\nu_{\rm obs}$ is:
\beqa
I(\nu_{\rm obs}, z=0) = \frac{1}{4 \pi} \int_0^\infty dz^\prime \frac{dl}{dz^\prime}\frac{\epsilon\left[\nu_{\rm obs} (1+z^\prime)\right]}{(1+z^\prime)^3}.
\label{eq:spec_intensity}
\eeqa
In this equation $dl/dz^\prime$ denotes the proper line element $dl/dz = c/\left[(1+z) H(z) \right]$.

Next we would like to compute the proper volume emissivity of the CO emitting sources. For simplicity, we assume that a (halo mass independent) 
fraction, $f_{\rm duty}$,
of halos with mass larger than $M_{\rm co, min}$ actively emit in the CO line of interest at any given time. Within the halos that host
CO luminous galaxies at a given time, we further assume that the CO luminosity is directly proportional to the halo mass.
Finally, we approximate the profile of each CO emission line as a delta function in frequency. The specific luminosity is then given by
\beqa
L_\nu = A \delta_D(\nu - \nu_J) M,
\label{eq:lum_mhalo}
\eeqa
where $\delta_D$ denotes a Dirac delta function, $\nu_J$ is the rest frame frequency of the transition of interest, $M$ is the mass
of the dark matter host halo, and $A$ is a proportionality constant that we will discuss shortly. 
Denoting the (co-moving) halo mass function at redshift $z$ by $dn_{\rm co}(z)/dM$, the proper volume emissivity is:
\beqa
\epsilon(\nu,z) = A \delta_D(\nu - \nu_J) (1+z)^3 f_{\rm duty} \times \nonumber \\ 
\int_{M_{\rm co, min}}^\infty dM \frac{dn_{\rm co}(z)}{dM} M.
\label{eq:emissivity}
\eeqa

Inserting Equations \ref{eq:lum_mhalo} and \ref{eq:emissivity} into Equation \ref{eq:spec_intensity} we find:
\beqa
I(\nu_{\rm obs}) = \frac{A}{4 \pi} \frac{1}{\nu_J} \frac{c}{H(z_J)} f_{\rm duty} \int_{M_{\rm co, min}}^\infty dM \frac{dn_{\rm co}(z_J)}{dM} M. \nonumber \\
\label{eq:intensity_model}
\eeqa
In this equation $\nu_{\rm obs} = \nu_J/(1 + z_J)$ is the observed frequency for gas at redshift $z_J$ emitting in a CO line with
rest frame frequency $\nu_J$. The equation implies that, in our simple model, the specific intensity of CO emission is directly proportional to the
fraction of matter in halos of mass greater than $M_{\rm co, min}$. 
We will generally quote results for the CO emission in terms of an equivalent 
brightness temperature, computed using the Rayleigh-Jeans Law, $T_{\rm CO}(\nu_{\rm obs}) = c^2 I(\nu_{\rm obs})/(2 k_B \nu_{\rm obs}^2)$.

\subsection{The Star Formation Model}
\label{sec:sfr_mod}

Next we need to specify the parameter $A$, which relates CO luminosity and host halo mass. The first ingredient involved in specifying $A$ is to connect
the star formation rate and host halo mass, $M$. Here we adopt the simplest plausible model and assume that the baryon fraction in the host halo
follows the universal value, $\Omega_b/\Omega_m$, and that a fraction $f_\star$ of these baryons are turned into stars at a constant rate over a time scale $t_s$. We assume
that $f_\star$ and $t_s$ are themselves independent of host halo mass. In this case
the SFR is (Loeb et al. 2005):
\beqa 
\rm{SFR} = && f_\star \frac{\Omega_b}{\Omega_m} \frac{M}{t_s} 
=  0.17 M_\odot \rm{yr}^{-1} \times \nonumber \\ 
&& \left[\frac{f_\star}{0.1}\right] \left[\frac{\Omega_b/\Omega_m}{0.17}\right] \left[\frac{10^8 \rm{yrs}}{t_s}\right] 
\left[\frac{M}{10^9 M_\odot}\right].
\label{eq:sfr_mhost}
\eeqa
We assume $f_\star = 0.1$ and $t_s = 10^8$ yrs. These values are consistent with those found by Stark et al. (2007) to match the luminosity function
of Lyman Break Galaxies (LBGs) at $z=6$ in a very similar model. 
Similarly, by comparing the UV luminosity function and the 2 point correlation function, Lee et al (2009) find the typical duration of star formation is $\lesssim 0.4$Gyr at $z \approx 4-6$. 
The star formation time scale $t_s$ sets the duty cycle for star
formation activity to be $t_s/t_H$; here $t_H$ is the Hubble time at the redshift of interest which is also approximately the `age'
of the dark matter halos.  The duty cycle for CO luminous activity, $f_{\rm duty}$, may differ from $t_s/t_H$ but is likely
comparable to this if CO is excited by starbursts (see \S \ref{sec:lum_co}). Our fiducial model assumes $f_{\rm duty} = 0.1$, which
is comparable to $t_s/t_H$ at the redshifts of interest.

It is also useful to consider the star-formation rate density (SFRD), $\dot{\rho}_\star(z)$, in our model. This depends on the collapse fraction of halos with mass larger than
$M_{\rm sf, min}$, where $M_{\rm sf, min}$ denotes the minimum dark matter halo mass that hosts a 
star-forming galaxy. 
Note that $M_{\rm sf, min}$ may be smaller than $M_{\rm co, min}$ if some galaxies
form stars yet do not emit in the CO line (see \S \ref{sec:lum_co}). The SFRD is then given by:
\beqa
\dot{\rho}_\star(z) = 0.079 M_\odot \rm{yr}^{-1}  \rm{Mpc}^{-3} \left[\frac{f_\star}{0.1}\right] \left[\frac{t_H(z=7)}{t_H(z)}\right] \times \nonumber \\ 
\left[\frac{f_{\rm coll}(M_{\rm sf, min},z)}{0.1}\right].
\label{eq:sfr_den}
\eeqa
The model star formation rate in the above equation depends on the Hubble time at the redshift of interest; the number above is scaled to the Hubble time at
$z=7$.  
Here $f_{\rm coll}(M_{\rm sf, min},z)$ denotes the fraction of mass in halos of mass greater than $M_{\rm sf, min}$ at redshift $z$; it is the halo `collapse fraction'.
The value of $10 \%$ for the collapse fraction in the above equation gives its order of magnitude at the typical redshifts and halo masses of interest.

Following Carilli (2011), it is useful to compare the SFRD in our model to the critical star formation rate density required
to overcome recombinations, and keep the Universe ionized (Madau et al. 1999). After reionization completes, the star formation rate density must clearly exceed this
critical value, while during most of the EoR one would expect the SFRD to be comparable to this critical rate. Hence although our star formation model is simplistic,
and the parameters in Equation \ref{eq:sfr_mhost} are uncertain, it is nevertheless constrained by
requiring the star formation density to be comparable to the critical value. If $\dot{\rho}_\star(z)$ were much less than (greater than) the critical rate, the Universe would be drastically
under-ionized (over-ionized) at the redshift of interest.

This critical SFRD depends on three uncertain quantities: the number of ionizing photons produced
per baryon converted into stars (which in turn depends on the stellar IMF and the spectral shapes
of the stars in the ionizing galaxies), the fraction of ionizing photons that escape from the host halo and ionize the IGM, and on
the clumpiness of the IGM. Adopting the ionizing spectrum expected for a Salpeter IMF, solar metallicity, and plausible values of the escape fraction and clumping
factor, the critical rate is given by (Madau et al. 1999, Munoz \& Loeb 2010):
\beqa
\dot{\rho}_{\star, \rm crit} = 0.050 M_\odot \rm{yr}^{-1} \rm{Mpc}^{-3} \left[\frac{1+z}{8}\right]^3 \left[\frac{0.1}{f_{\rm esc}}\right] \left[\frac{C}{5}\right].
\label{eq:sfr_crit}
\eeqa
The above choice of clumping factor, $C=5$, is similar to that measured from recent hydrodynamical simulations (e.g. Pawlik et al. 2009, McQuinn et al. 2011),
although there are still theoretical uncertainties in this estimate. In particular, note that the clumping factor that enters 
Equation \ref{eq:sfr_crit} is that of {\em the ionized gas in the IGM}. The clumping factor then depends on which gas is 
considered to be part of the IGM, and which gas belongs to a galaxy, as well as the properties of reionization itself, such
as the spatial distribution and spectrum of the ionizing sources. The value of the escape fraction in the 
above equation, $f_{\rm esc} =0.1$,
is similar to that measured from LBGs at $z \sim 3$ (Nestor et al. 2011). However, the structure of
the faint dwarf galaxies that likely reionized the Universe may be rather different than that of these LBGs, and the reionizing 
galaxies may hence have rather different escape fractions than the $z \sim 3$ LBGs. While the uncertainties are substantial, Equation
\ref{eq:sfr_crit} still provides a useful indication of plausible values for the high redshift star formation rate. 

The claim that the SFRD should not differ greatly from this
critical value at the redshifts of interest is strengthened by considering measurements of the mean transmitted flux
in the Ly-$\alpha$ forest at high redshift. These suggest 
that the total emissivity of ionizing photons
at $z=5-6$ is comparable to the critical value required to keep the Universe
ionized (Miralda-Escud\`e 2003, Bolton \& Haehnelt 2005, Faucher-Gigu\`ere et al. 2008, McQuinn et al. 2011).  In addition to the Ly-$\alpha$ forest measurements at $z \sim 5-6$, 
WMAP constraints on the electron scattering optical depth (Komatsu et al. 2010) suggest significant amounts of star formation at $z \gtrsim 6$; it is hence unlikely that the
SFRD falls steeply right above $z \sim 5-6$.
The Ly-$\alpha$ forest data indicate a hydrogen photoionization rate of $\Gamma_{\rm HI} \approx 5 \times 10^{-13} s^{-1}$ at $z=5.5$ (see the above papers for a discussion of the uncertainties in this measurement). Combining this with a recent constraint on
the mean free path to HI ionizing photons, $\lambda_{\rm HI}$, from Songaila \& Cowie (2010), and assuming that
the ionizing spectrum is a $\nu^{-2}$ power law near the HI photoionization edge gives:
\beqa
\dot{\rho}_{\star, Ly-\alpha} = 0.039 M_\odot \rm{yr}^{-1} \rm{Mpc}^{-3} \left[\frac{\Gamma_{\rm HI}}{5 \times 10^{-13} s^{-1}}\right] \times \nonumber \\ 
\left[\frac{10^{53.1} s^{-1}}
{N_{\rm ph}}\right] \left[\frac{0.1}{f_{\rm esc}}\right] \left[\frac{9.8 p \rm{Mpc}}{\lambda_{\rm HI}}\right] \left[\frac{6.5}{1+z}\right]^3.
\label{eq:sfr_lya}
\eeqa
In this equation, $N_{\rm ph}$ is the number of ionizing photons emitted per second for an
$\rm{SFR}$ of $1 M_\odot$ yr$^{-1}$, and
$10^{53.1} s^{-1}$ is the value expected for a Salpeter IMF and solar metallicity (Madau et al. 1999).\footnote{For simplicity of presentation, we do not include separate bracketed terms illustrating an additional (weak) dependence on the assumed shape of the ionizing spectrum.} The number in brackets for $\lambda_{\rm HI}$ is the preferred value of the mean free path in units of proper Mpc from Songaila \& Cowie (2010). Taken together, Equations \ref{eq:sfr_crit}
and \ref{eq:sfr_lya} argue that $\dot{\rho}_\star(z)$ is unlikely to differ significantly
from $\dot{\rho}_{\star,crit}$ for most of the redshifts of interest, except perhaps in the
earliest phases of the EoR, at which point the CO measurement is likely impossible anyway. 
Further,
the equations quantify the expected SFRD, which is $\dot{\rho}_\star(z = 7) \sim 0.05 M_\odot$ yr$^{-1}$ Mpc$^{-3}$, although with
significant uncertainties from the unknown value of $C/f_{\rm esc}$.  

Finally, we compare our model SFRD with these numbers. The model depends
on $M_{\rm sf, min}$, the minimum mass halo hosting a star-forming galaxy. One plausible value
for $M_{\rm sf, min}$ is the halo mass corresponding to a virial temperature of 
$T_{\rm vir} = 10^4 K$ -- $M_{\rm sf, min} \approx 10^8 M_\odot$ at the redshifts of interest -- above which gas is able to cool by atomic hydrogen line emission,
condense, and form stars (e.g. Barkana \& Loeb 2001). This mass scale is sometimes called the `atomic cooling mass'. For $M_{\rm sf, min} = 10^8 M_\odot$, we compute the model SFRD of Equation \ref{eq:sfr_den} at $z=7$ using
the Sheth-Tormen (2001) halo mass function, finding 
$\dot{\rho}_\star(z=7) = 0.060 M_\odot$ yr$^{-1}$ Mpc$^{-3}$. This is indeed comparable
to the critical rate (Equations \ref{eq:sfr_crit}, \ref{eq:sfr_lya}).
Feedback from
photoionization and supernovae explosions may, however, limit the efficiency of star formation in these small mass halos, raising $M_{\rm sf, min}$, although the precise impact of these
effects is still unclear (e.g. Dijkstra et al. 2004). If $M_{\rm sf, min} = 10^9 M_\odot$, for example, the model SFRD drops to $\dot{\rho}_\star(z=7) = 0.029 M_\odot$ yr$^{-1}$ Mpc$^{-3}$, just a little lower than the critical rate for $f_{\rm esc} = 0.1$, $C=5$ (Equation \ref{eq:sfr_crit}).
The SFRD in this model is closer to the critical rate if the escape fraction is higher, the clumping factor lower, or
if the normalization of the SFR-$M$ relation is higher (Equation \ref{eq:sfr_mhost}).
With the
latter solution, the SFRD would again be similar to that in our atomic cooling mass model. In other words, normalizing the model to the 
critical SFRD removes at least some of the freedom from
varying $M_{\rm sf, min}$ (see also Carilli 2011, and the present paper, \S \ref{sec:param_var_auto}). 
Note also that $M_{\rm sf, min}$ might be smaller than the atomic cooling mass 
if molecular hydrogen cooling is efficient despite negative feedback from dissociating UV radiation (Haiman et al. 1997). Given these and other uncertainties,
we will simply use the atomic cooling mass as our
fiducial $M_{\rm sf, min}$, along with the SFR-$M$ relation of Equation \ref{eq:sfr_mhost}, since this model produces approximately the expected SFRD.

\subsection{CO Luminosity}
\label{sec:lum_co}

The next crucial, but least certain, ingredient in our model is to relate the CO luminosity and star formation rate at high redshift. 
We will consider this from both an empirical point of view and from
a theoretical perspective. Observationally, CO emission lines have been detected all the way out to $z \sim 6$ from the host galaxies of bright quasars, providing part of the impetus
for our present study (Walter et al. 2003, Wang et al. 2010). Similarly, CO emission lines have been detected at high redshift ($z \sim 4.5$) from submilimeter galaxies with
extreme starbursts ($SFR \sim 10^3 M_\odot$ yr$^{-1}$) (e.g. Schinnerer et al. 2008). Much less clear observationally is whether more normal galaxy populations,
without extreme starbursts, are CO luminous at high redshift: so far there are only upper limits on CO(1-0) emission from a pair of $z = 6.6$ LAEs (Wagg et al. 2009). 
From a theoretical perspective, the CO emission depends on complex astrophysical processes that are difficult to model from first principles. For example, it depends on the structure of high 
redshift galaxies and the spatial
distribution of their star forming gas, and the metallicity and temperature of the molecular clouds in these galaxies. The 
temperature of the molecular clouds in turn depends on the level of heating
from the CMB, photoelectric heating by dust grains and heating from  starbursts, AGN activity, supernova shocks, and cosmic rays, as well as cooling by atomic fine structure lines and molecular lines (e.g. Obreschkow et al. 2009).

Indeed, Obreschkow et al. (2009) present a detailed model for CO emission from high redshift galaxies, incorporating models for many of these physical processes. Their model starts from
a semi-analytic model of galaxy formation placed on top of the Millenium simulation (Springel et al. 2005), and then incorporates additional modeling to split the cold gas into atomic and molecular components and
for converting H$_2$ abundances into CO luminosity. One approach would to use the Obreschkow et al. (2009) model directly, as done in Gong et al. (2011). While we draw heavily on the discussion
in Obreschkow et al. (2009) to inform our modeling, we prefer a different approach for several reasons. First, the minimum resolved halo mass in the Millenium simulation is $3 \times 10^{10} M_\odot$ (
Springel et al. 2005),
more than two orders of magnitude larger than the atomic cooling mass; hence the hosts of the likely ionizing sources are not resolved in this calculation. 
Furthermore, the models for these massive galaxies make assumptions: e.g., that all galaxies are quiescent, virialized exponential disks---which drives the partition between ${\rm H_{2}}$ and HI---which is unlikely to apply in the small, constantly merging systems which drive the CO signal at high redshift. 
Next, observed galaxy properties
at low and intermediate redshift are relatively well understood --- and the semi-analytic models are well-calibrated at these redshifts -- but the properties of low luminosity galaxies at $z \gtrsim 6$
are much less certain. 
The use of a highly complex, multi-parameter model with potentially inapplicable physics obfuscates the key requirements--and associated uncertainties---for a detectable signal. 
As a result, we will adopt a much simpler model for CO luminosity and try to identify key sources of uncertainty.

Let us start by considering empirical correlations between star formation rate, far-infrared luminosity, and CO luminosity as measured in relatively nearby galaxies (see also Carilli 2011).
Wang et al. (2010) fit a correlation between far-infrared luminosity and CO luminosity using measurements from galaxies at $z=0-3$, including spiral 
galaxies, LIRGs, ULIRGs, and SMGs. Specifically they
fit a power law relation between far-infrared luminosity and the velocity integrated CO(1-0) brightness temperature multiplied by the source area. The latter quantity is denoted by
$L^\prime_{\rm CO(1-0)}$, and has dimensions of K km/s pc$^2$. It is related to the total luminosity of the emission line, $L_{\rm CO(1-0)}$, by the relation:
\beqa
L_{\rm CO(1-0)} = 1.04 \times 10^{-3} L_\odot \left[\frac{L^\prime_{\rm CO(1-0)}}{3.25 \times 10^7 \rm{K \, km \,} \rm{s}^{-1}   \rm{\, pc}^2}\right]. \nonumber \\
\label{eq:lprime_def}
\eeqa 
The Wang et al. correlation is:
\beqa
\rm{log}\left(\frac{L_{\rm FIR}}{L_\odot}\right) = 1.67 \rm{log}\left(\frac{L^\prime_{\rm CO(1-0)}}{\rm{K \, km \,} \rm{s}^{-1} \rm{\, pc}^2}\right) 
- 4.87,
\label{eq:wang_corr}
\eeqa
with $L_{\rm FIR}$ denoting the far-infrared luminosity.
This can then be combined with the Kennicutt (1998) correlation between SFR and $L_{\rm FIR}$, which is $\rm{SFR} = 1.5 \times 10^{-10} M_{\odot} \rm{yr}^{-1} (L_{\rm FIR}/L_\odot).$
Assuming this relation between far-infrared luminosity and SFR, the Wang et al. (2010) fit covers a wide range of SFRs from $\sim 0.1-1,000 M_{\odot} $ yr$^{-1}$.
Combining the Kennicutt relation with Equations \ref{eq:lprime_def} and \ref{eq:wang_corr}, we arrive at a relationship between the luminosity in the CO(1-0) line, and the star formation rate:
\beqa
L_{\rm CO(1-0)} = 3.2 \times 10^4 L_\odot \left[\frac{\rm{SFR}}{M_\odot \rm{yr}^{-1}}\right]^{3/5}.
\label{eq:lum_co}
\eeqa
Note that according to this relation the CO luminosity is a sub-linear function of the SFR. This acts to weight the contribution  
of galaxies with low star formation rates to the emissivity (Equation \ref{eq:emissivity}) most heavily. This implies that CO emission may be especially bright
if the scaling continues to apply at high redshifts ($z \gtrsim 6$)  and low SFRs ($SFR \lesssim 0.1 M_\odot$ yr$^{-1}$), since abundant dwarf galaxies with low SFRs are likely ionizing
sources.

One possible theoretical explanation for the sub-linear scaling of CO luminosity with SFR
was proposed by Narayanan et al. (2010): the CO(1-0) transition has a low critical density (to thermalize) and so emission in this line may be independent of SFR and simply proportional to the
gas density, $L_{\rm CO(1-0)} \propto \rho$. Combining this with the Schmidt (1959) law, $SFR \propto \rho^{3/2}$, gives approximately 
the observed sub-linear scaling. If this explanation is correct, it may hold at higher redshift as well although as discussed below, a sufficiently
strong starburst is required to produce a high enough excitation temperature, $T_{\rm ex}$, for CO to be observable
against the high redshift CMB.
Lines from significantly
higher rotational states will have higher critical densities, and the excitation rate of these lines will then depend on the SFR, and give a different, stronger scaling of $L_{\rm CO}$ w/ SFR, as
found observationally (Naryanan et al. 2010). These authors' simulations give, however, a similar sub-linear 
scaling for CO(2-1) and CO(1-0), and so this scaling appears reasonable for the two
emission lines of interest for our study.

Let us now consider the various effects that may modify the empirical CO luminosity-SFR relation of Equation \ref{eq:lum_co} at high redshift. 
One significant worry is that the CO emission must be detected {\em against the CMB}, whose temperature grows
with redshift as $T_{\rm CMB} \propto (1+z)$. In the limit that the CO excitation temperature, $T_{\rm ex}$, is equal to the CMB temperature, the CO emission is completely 
{\em undetectable}
at the redshift of interest (Combes et al. 1999, Obreschkow et al. 2009).  
As mentioned in the beginning of this section, a number of processes may be important in
determining the CO excitation temperature, and it is unclear how these will impact the molecular gas in the high 
redshift dwarf galaxies of interest. In particular, although
high redshift dwarfs are small in size and have low luminosities, the relevant quantity for computing their excitation temperature is the star formation density. 
Specifically, in the optically thick limit, $T_{\rm ex}^{4} \approx T_{\rm CMB}^{4} + T_{\rm heat}^{4}$, where $T_{\rm heat}$ reflects the dust heating due to starburst and/or AGN activity; in order for the CMB not to significantly dilute the CO emission, $T_{\rm heat} \ge T_{\rm CMB}$. 
The molecular gas is likely very
dense in these high redshift galaxies, and so their star formation densities should be large, provided their star formation obeys a Schmidt-type law. Note that in starbursts, $T_{\rm ex} \ge 40$K in observed CO emitters (Obreschkow et al. 2009), which still gives plenty of
contrast with the CMB at $z \sim 8$, for example. 
Indeed, 
given the potentially higher densities in high redshift galaxies, 
it is possible
that high redshift dwarf galaxies are more luminous in CO than expected from the low redshift empirical constraints
(although note that this is more likely to boost detectability in high J transitions). 

A second possible worry is that the metallicity will decrease towards high redshift and that this will rapidly diminish the CO emissivity at early cosmic times.
This is likely less severe a problem than it might first appear. First, while the spatial average metallicity is likely quite low at $z \gtrsim 6$, rapidly star-forming regions
will typically already be enriched by several generations of stars. Second, the low order CO transitions likely come from optically thick regions, in which case the emission
may be nearly independent of metallicity. However, there is likely less dust in these high redshift star-forming regions than in their low redshift counterparts, and the dust helps
shield CO from UV dissociating radiation (e.g. Obreschkow et al. 2009). 

Given the uncertainties, we will stick to the simplest plausible model outlined in \S \ref{sec:co_bright}. This model is parametrized
by $M_{\rm co, min}$, the minimum host halo mass for CO luminous halos; the luminosity of such halos, $L_{\rm CO} (M_{\rm co, min})$ (with $L_{\rm CO} \propto M$ for larger mass halos);
and the duty cycle, $f_{\rm duty}$, for CO activity. The latter parameter is entirely degenerate with $L_{\rm CO} (M_{\rm co, min})$ as far as the mean specific intensity of
the CO emission is concerned (Equation \ref{eq:intensity_model}), but the duty cycle further impacts the level of Poisson fluctuations in the CO emission (\S \ref{sec:co_flucs}).  
The parameter $M_{\rm co, min}$ will crudely mimic the scenarios discussed above in which low SFR galaxies are dim in the CO lines of interest at high redshift.
For instance, $M_{\rm co, min}$ may be large if dwarf galaxies with low SFRs do not boost the gas temperature and the 
excitation temperature of the low-J CO lines, $T_{\rm ex}$, enough to be visible
above the CMB background at $z \gtrsim 6$, or if the metallicity of the star-forming regions in these galaxies is too low and CO is dissociated as a result.  

{\em Our aim is to understand which values of these parameters produce a detectable CO power spectrum and 21 cm-CO cross power spectrum signal.}
In our fiducial calculations, we assume $M_{\rm co, min} = M_{\rm cool} \sim 10^8 M_\odot$, $f_{\rm duty} = 0.1$. We further adopt the (lower redshift) 
empirical CO luminosity-SFR relation of Equation \ref{eq:lum_co}. For simplicity, we take $L_{\rm CO} \propto M \propto SFR$ and so our model will not precisely match the
sub-linear scaling of Equation \ref{eq:lum_co} for all model galaxies. However, we normalize the model CO luminosity to match
that demanded by the {\em sub-linear} scaling of Equation \ref{eq:lum_co} for star forming halos with the atomic cooling mass.  In other words, we fix $L_{\rm CO} (M= M_{\rm cool})$
using Equation \ref{eq:lum_co} and assume $L_{\rm CO} \propto M$ for larger mass halos. In \S \ref{sec:param_var_auto} we check the impact of relaxing this assumption,
and generalizing the dependence to $L_{\rm CO} \propto M^\alpha$. There we find that assuming the linear scaling for masses above
$M_{\rm cool}$, rather than the sub-linear one, has little impact on our results. This occurs because the emissivity in 
these models is dominated
by the sources in the smallest halos. For instance, the mean emissivity is only a factor of two higher in the linear case.

Combining Equation \ref{eq:sfr_mhost} and Equation \ref{eq:lum_co}, we obtain
$L_{\rm CO} = 2.8 \times 10^3 L_\odot$ for halos of mass $M=10^8 M_\odot$.
Our fiducial model then assumes
\beqa
L_{\rm CO}(M) = 2.8 \times 10^3 L_\odot \left[\frac{M}{10^8 M_\odot}\right] .
\label{eq:lumfid_mhalo}
\eeqa
The empirical relation (Equation \ref{eq:lum_co}) is for the CO(1-0) line, and so the above equation strictly applies only to it, and not to emission lines from higher rotational
levels. However, given the uncertainties
in the normalization of the CO luminosity at high redshift, we will assume that it applies to the CO(2-1) line as well, which is very 
likely a conservative assumption. For example, in the limit that both lines are optically thick and $k_B T_{\rm ex} >> h \nu_J$,  $L_{\rm CO(2-1)} = 8 L_{\rm CO(1-0)}$ (e.g. Obreschkow et al. 2009), although the ratio of the luminosities in the two lines is unlikely as large
as in this limit.\footnote{For reference, $h \nu_J/k_B = J 5.5 K$.}
The relation in Equation \ref{eq:lumfid_mhalo} then fixes our fiducial choice of the parameter $A$ in Equations \ref{eq:lum_mhalo}, \ref{eq:emissivity}, and \ref{eq:intensity_model}.

For the CO(2-1) line, the normalization of Equation \ref{eq:lumfid_mhalo} is a factor of $6$ higher than that adopted in 
Righi et al. (2008) at the star formation rate of atomic cooling mass halos in our model. Our CO(1-0) luminosity
is higher than theirs by a still larger factor. 
Their lower normalization is driven by their matching to M82, which has an SFR
of $\sim 10 M_\odot$ yr$^{-1}$. This is significantly larger than the expected SFRs of the low luminosity dwarf galaxies that likely reionized the Universe, and so we
prefer our normalization which may provide a 
better match to the low luminosity galaxies of interest. In \S \ref{sec:previous} we provide
further comparison with previous work.

\section{Spatially-Averaged CO Emission}
\label{sec:mean_co}

We now have all of the basic ingredients necessary to calculate CO brightness temperatures in our model. Combining Equations \ref{eq:intensity_model}, \ref{eq:sfr_mhost}, and \ref{eq:lumfid_mhalo} with the Rayleigh-Jeans relation between brightness temperature and specific intensity, and inserting our fiducial model parameters, we obtain:
\beqa
T_{\rm CO}\left(\nu_{\rm obs}\right) = 2.1 \mu K \left[\frac{f_{\rm coll}(M_{\rm co, min}; z_J)}{0.1}\right] \left[\frac{f_{\rm duty}}{0.1}\right]  \times \nonumber \\
\left[\frac{2}{J}\right]^3 
\left[\frac{1+z_J}{8}\right]^{1/2}.\nonumber \\
\label{eq:tco_fid}
\eeqa
In this equation, gas emitting at a redshift $z_J$ in a CO transition of rest frame emission frequency $\nu_J$ is observed at a frequency $\nu_{\rm obs} = \nu_J/(1+z_J)$. 
The frequency emitted in a transition between rotational levels $J \rightarrow J-1$ is $\nu_J = J \nu_{\rm CO}$, with $\nu_{\rm CO} = 115$ Ghz.
The fiducial value for $J$ in the above equation is $J=2$, i.e. the numbers are for the CO(2-1) transition.  
CO gas at the same redshift emitting in the CO(1-0) transition will
be observed at half of the frequency of the CO(2-1) line. This equation makes the conservative assumption that the CO(2-1) line
and the CO(1-0) line have the same luminosity-halo mass relation. In the high temperature and optically thick limit mentioned in the
previous section, and adopting Equation \ref{eq:lumfid_mhalo} for the CO(1-0) line, the brightness temperature in 
the CO(2-1) line would be a factor of $8$ higher than implied by Equation \ref{eq:tco_fid}.
For simplicity, we will focus most of the remaining discussion
on the CO(2-1) line and adopt Equation \ref{eq:tco_fid}: the reader can rescale the results for the CO(1-0) line based on Equation \ref{eq:tco_fid} and their preferred luminosity-halo mass relation.
This compact equation then allows us to predict the CO brightness temperature for a region in which a fraction $f_{\rm coll}$  of matter has collapsed into dark matter halos
hosting CO luminous galaxies.
Note that $T_{\rm CO}$ and $f_{\rm coll}$ will fluctuate from place-to-place across the Universe, but we suppress the spatial coordinate in our notation here.

\begin{figure}
\bc
\includegraphics[width=9.2cm]{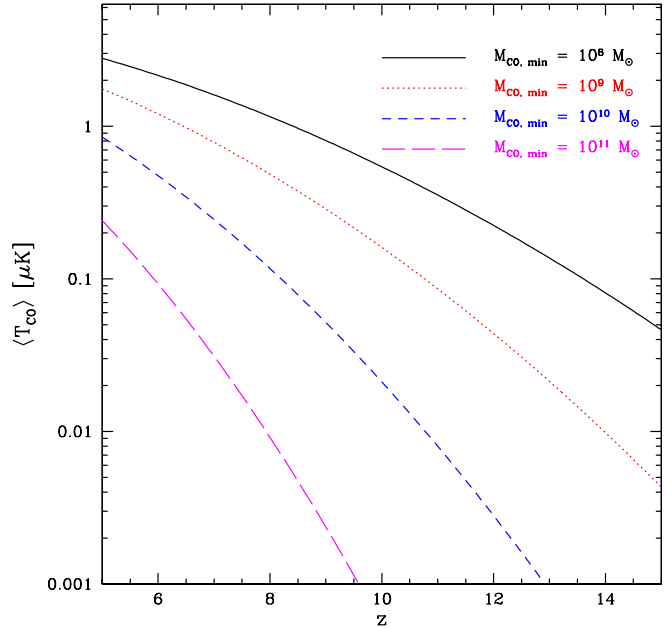}
\caption{The global mean CO brightness temperature as a function of
redshift.  The curves show the volume-averaged CO brightness temperature model of Equation \ref{eq:tco_fid} for several different values of $M_{\rm co, min}$, the minimum
halo mass hosting a CO luminous galaxy. The models assume that the minimum host halo mass of star forming galaxies is the atomic cooling mass, $M_{\rm sf, min} \approx 10^8 M_\odot$,
and so the curves with larger $M_{\rm co, min}$ describe models in which galaxies with low SFRs are not CO luminous. If we had instead varied $M_{\rm sf, min}$, while fixing both
$M_{\rm co, min} = M_{\rm sf, min}$ and the SFRD at a given redshift, the model variations would be considerably smaller (see text).}
\label{fig:tco_mean}
\ec
\end{figure}
We start by computing the volume-averaged CO brightness temperature. In order to illustrate some of the dependence on model uncertainties, we vary
the minimum halo mass hosting a CO luminous galaxy widely, between $M_{\rm co, min} = 10^8, 10^9, 10^{10}$, and $10^{11} M_\odot$. In our model, these halos host star formation rates of
$\sim 0.02, 0.2, 2,$ and $20 M_\odot$ yr$^{-1}$, respectively.  
Note that this minimum mass may be distinct from
the minimum mass dark matter halo in which stars form (we denote this latter mass by $M_{\rm sf, min}$) if some galaxies are simply not CO luminous, as discussed in the previous section. 

The results of the mean brightness temperature calculation are shown in Figure \ref{fig:tco_mean}. For this calculation, we use the Sheth-Tormen (2001) mass function to calculate
the collapse fraction, since this formula matches the abundance of simulated dark matter halos at high redshifts 
fairly well (e.g. Zahn et al. 2007). In all models, the mean CO brightness temperature
is a steep function of redshift, which results because the host halos are rare objects at the initial redshifts of interest, and the collapse fraction consequently grows rapidly towards lower
redshifts.  This steep redshift evolution is enhanced for the large values of $M_{\rm co,min}$, 
in which case the CO luminous host halos are correspondingly rare. 
At $z \sim 7$ the mean brightness temperature is of order $1 \mu K$ for $M_{\rm co, min} = 10^8 M_\odot$, but the results fall with increasing $M_{\rm co, min}$. For example, the mean brightness temperature at $z \sim 7$ is a factor of $\approx 2$ smaller than this
 if $M_{\rm co, min} = 10^{9} M_\odot$, and
it is a factor of $\sim 7$ smaller than this if $M_{\rm co, min}$ is as large as $10^{10} M_\odot$ (SFR $= 2 M_\odot$ yr$^{-1}$ in our model). The mean emission drops off still further
in the extreme case that $M_{\rm co, min} = 10^{11} M_\odot$, in which case $\avg{T_{\rm CO}}$ is about a factor of $50$ smaller than in the atomic cooling mass model.  
The strong dependence on $M_{\rm co, min}$ arises because the collapse
fraction depends strongly on the minimum host halo mass for these rare halos.
At still higher redshifts, the relevant
dark matter halos are rarer, and the dependence on $M_{\rm co, min}$ is still stronger.  As we will detail subsequently, the dependence shown here has potentially 
strong implications for the feasibility of CO intensity mapping
experiments at high redshift. 

It is important to emphasize, however, that the most important source of uncertainty is not so much in the minimum mass of halos that host star-forming galaxies ($M_{\rm sf, min}$), but rather in the minimum host halo mass of {\em CO luminous} galaxies ($M_{\rm co, min}$). This is because, regardless of $M_{\rm sf, min}$, the SFRD must be comparable to the critical rate (Carilli 2011) once reionization
is significantly underway. In the context of our model, boosting $M_{\rm sf, min}$ necessitates increasing the SFR-M normalization 
to ensure that the SFRD is still comparable
to the critical rate. In our fiducial model, both $\avg{T_{\rm CO}}$ and the SFRD are proportional to the collapse fraction, with $\avg{T_{\rm CO}}$ scaling with
the SFR-M normalization to the $3/5$th power (because $L_{\rm CO}$ scales sub-linearly with SFR). Hence normalizing to a fixed SFRD at a given redshift will remove
most of the dependence on $M_{\rm sf, min}$ at that redshift. We quantify this further in \S \ref{sec:param_var_auto}.
The only caveat with this argument is that the precise normalization of the SFRD is subject to considerable uncertainties (\S \ref{sec:sfr_mod}). 
Nonetheless, the main concern is not in the value of $M_{\rm sf, min}$ but that the galaxies that produce most of the ionizing photons -- likely abundant sources with low star formation rates -- 
may be CO dim for the reasons discussed in \S \ref{sec:lum_co}. 
The value of $M_{\rm sf, min}$ does of course impact the strength of brightness temperature fluctuations, as we discuss in \S \ref{sec:co_flucs}. 

Finally, let us discuss how the spatial mean brightness temperature depends on the duty cycle for CO luminous activity, $f_{\rm duty}$.
First note that the SFRD is independent of the starburst duty cycle: as the star formation timescale increases, an increasing
fraction of gas needs to be converted into stars to produce a given SFRD (Equations \ref{eq:sfr_mhost}-\ref{eq:sfr_den}). 
Because of this, the duty cycle impacts $\avg{T_{\rm CO}}$ most strongly if the duty cycle for CO luminous activity is significantly
larger or smaller than that for star formation. Since CO lines are excited by starburst activity, we generally expect the two duty
cycles to be comparable, as assumed in our fiducial model. If the CO duty cycle is instead larger (smaller) than
this fiducial starburst value, the mean brightness temperature will be increased (decreased) in direct proportion.

\section{CO Brightness Temperature Fluctuations}
\label{sec:co_flucs}

Unfortunately, it is likely impossible to observe the global mean CO brightness
temperature directly.
The difficulty with observing this signal
is that there will be other sources of emission at the relevant frequencies that are much brighter than the redshifted CO signal. 
However, similar to the case
of the redshifted 21 cm signal, the CO signal should have structure in frequency space while foreground sources will be spectrally 
smooth (see \S \ref{sec:trans_co}). One can use this
fact to isolate spatial fluctuations in the CO brightness temperature, and detect them robustly.
 
\subsection{Numerical Simulations}
\label{sec:num_sims}

In order to model spatial fluctuations in the CO emission and the redshifted 21 cm signal, we use outputs from the numerical simulations of reionization
of McQuinn et al. (2007a, 2007b). These simulations treat the radiative transfer of ionizing photons in a post-processing step performed on top of an evolved N-body simulation.
The N-body simulation was run using an enhanced version of Gadget-2 (Springel 2005), and follows $1024^3$ dark matter particles in a box with a co-moving
side length of $130 \hmpc$.  The simulation directly resolves halos down to $\sim 10^{10} M_\odot$, but smaller mass halos down to the atomic cooling mass scale
of $M_{\rm cool} \sim 10^8 M_\odot$ are incorporated with the appropriate abundance and clustering as in McQuinn et al. (2007a).
This is done using a merger tree algorithm similar to the PThalo code (Scoccimarro \& Sheth 2002), a widely used code for rapidly
generating mock galaxy surveys.  
In the previous section we found that halos
close to the atomic cooling mass, if indeed CO luminous, may provide the dominant contribution to the spatially-averaged CO brightness temperature and so it 
is important
to accurately capture the impact of these halos. Note that for the CO calculations we only require simulated halo catalogs from 
the N-body simulation, and not the full reionization
simulations. To calculate the 21 cm field, and the 21 cm-CO cross power spectrum, we use reionization calculations from the same fiducial 
model as in Lidz et al. (2008). In
this model, a source's ionizing luminosity is directly proportional to its host halo mass for halos above the atomic cooling mass.

\subsection{The CO Auto Power Spectrum}
\label{sec:co_auto}

In order to characterize the spatial fluctuations in the CO emission, we focus on
the power spectrum of CO brightness temperature fluctuations. We construct (three-dimensional) maps of
CO emission from simulated halo catalogs using Equation \ref{eq:tco_fid} for various
redshift outputs and values of $M_{\rm co, min}$. For any given redshift output, we randomly
select a fraction $f_{\rm duty}$ of the simulated halos to be CO-luminous, and assume that the remaining
fraction, $1 - f_{\rm duty}$, of the halo catalog are not actively emitting in CO.
We then measure the spherically-averaged auto power spectrum of each CO map using Fast-Fourier
Transforms (FFTs). We ignore the impact of peculiar velocities which have a small impact on the spherically averaged power spectra
compared to the uncertainties in the model. 

\begin{figure}
\bc
\includegraphics[width=9.2cm]{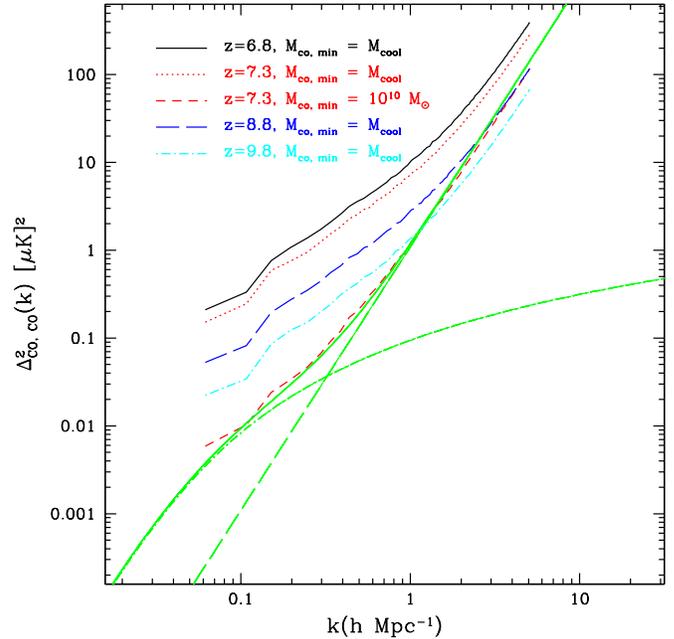}
\caption{The auto power spectrum of CO brightness temperature fluctuations. The black solid,
red dotted, red dashed, blue dashed, and cyan dot-dashed curves show simulated CO power spectra
at different redshifts for various values of $M_{\rm co, min}$. In each case the duty cycle is
fixed at $f_{\rm duty}=0.1$. The green solid line is the halo model of Equation \ref{eq:power_mod} 
for $z=7.3$, $M_{\rm co, min} = 10^{10} M_\odot$, and $f_{\rm duty} = 0.1$. The green dashed line 
shows the Poisson term, while the green dot-dashed curve is the clustering term.
}
\label{fig:tco_auto}
\ec
\end{figure}

Power spectra for several simulated models are shown in Figure \ref{fig:tco_auto}.
The figure shows $\Delta^2_{\rm CO, CO}(k)=k^3 P_{\rm CO, CO}(k)/(2 \pi^2)$, which denotes the contribution
to the variance of the CO brightness temperature per ln$(k)$, in units of $(\mu K)^2$,
as a function of $k$.
The strength of the CO brightness temperature fluctuations evidently depends
strongly on the model, redshift, and spatial scale. At a wavenumber of $k = 0.1 h$ Mpc$^{-1}$ and
at $z=6.8$, the amplitude of the power spectrum is  $\Delta^2_{\rm CO,CO} = 0.2 (\mu K)^2$ for
$M_{\rm co, min} = 10^8 M_\odot$,
while the power spectrum is down by about an order of magnitude on this scale by $z=9.8$.
The results appear quite sensitive to $M_{\rm co, min}$; in the model
with $M_{\rm co, min} = 10^{10} M_\odot$, the power spectrum at $k = 0.1 h$ Mpc$^{-1}$ is a factor of $\sim 20$
smaller than in the model with $M_{\rm co, min} = 10^8 M_\odot$. 
However, if we vary $M_{\rm sf, min}$ while fixing the SFRD (by adjusting the relationship between halo mass and SFR)
and fixing $M_{\rm co, min} = M_{\rm sf, min}$, this sensitivity is vastly reduced (\S \ref{sec:param_var_auto}).  

The fluctuations also
depend strongly on wavenumber, with the fluctuations typically reaching $\Delta^2_{\rm CO,CO}(k) \sim 100 (\mu K)^2$
on scales of $k \sim$ a few $h$ Mpc$^{-1}$.

We can understand all of these trends quantitatively using a `halo model' (Cooray \& Sheth 2002)
type calculation. In particular, we expect the auto power spectrum of the CO brightness temperature fluctuations to have the
following form:
\beqa
P_{\rm CO,CO}(k) = \avg{T_{\rm CO}}^2 \left[\avg{b}^2 P_{\rm lin}(k) + \frac{1}{f_{\rm duty}}\frac{\avg{M^2}} {\avg{M}^2}\right].
\label{eq:power_mod}
\eeqa
In this equation $P_{\rm lin}(k)$ denotes the linear theory density power spectrum, while 
\beqa
\avg{b} = \frac{\int_{M_{\rm co, min}}^\infty dM M \frac{dn_{\rm co}}{dM} b(M)}{\int_{M_{\rm co, min}}^\infty dM M \frac{dn_{\rm co}}{dM}}
\label{eq:bbar}
\eeqa
is a mass-weighted bias, which enters here because the model CO emission is proportional
to host halo mass. The second term in Equation \ref{eq:power_mod} involves the mean and
second-moment of the halo mass function,
\beqa
\avg{M^2} = \int_{M_{\rm co, min}}^\infty dM M^2 \frac{dn_{\rm co}}{dM} ; \nonumber \\
\avg{M} = \int_{M_{\rm co, min}}^\infty dM M \frac{dn_{\rm co}}{dM},
\label{eq:apoiss}
\eeqa
and represent the shot noise contribution to the power spectrum. These equations assume: our usual model for connecting CO luminosity and host halo mass, 
linear biasing, that the scale of interest is much larger than the
virial radius of the relevant halos, and that the halo shot noise obeys Poisson statistics. 
The virial radius of
a $10^8 M_\odot$ halo at $z=7$ is $15$ co-moving kpc (Barkana \& Loeb 2001), and so neglecting the impact of the halo
profile should be a good approximation on the relevant scales. 
The formula also assumes that CO active halos host precisely one CO luminous galaxy located at the halo center -- as does our simulated model -- likely 
a good assumption for the low mass halos of interest.
In order to calculate the terms in Equations \ref{eq:power_mod} -- \ref{eq:apoiss}, we
use the Eisenstein \& Hu (1999) transfer function, which was used to generate the initial conditions of the McQuinn et al. (2007b) reionization simulations, the Sheth-Tormen (2002) mass function, and
the Sheth et al. (2001) formula for halo bias.

Most of the power spectrum's redshift evolution, and its variation with $M_{\rm co, min}$, is driven by
its quadratic dependence on $\avg{T_{\rm CO}}$. $\avg{T_{\rm CO}}$ is in turn proportional to the halo collapse fraction in our model 
(Equation \ref{eq:tco_fid}).
This is partly compensated by $\avg{b}^2$, which increases as the
host halos become rarer and hence more clustered. The decrease in $\avg{T_{\rm CO}}$ is
the dominant effect, however, as one can see by comparing the $M_{\rm co, min} =
M_{\rm cool}$ and $10^{10} M_\odot$ curves. The form of Equation \ref{eq:power_mod}
also explains the shape of the simulated power spectra in Figure \ref{fig:tco_auto}.
On large scales the clustering term dominates and the CO fluctuations trace the underlying density power spectrum, while on small scales the Poisson term is
the main source of fluctuations. The green lines in Figure \ref{fig:tco_auto} show
that the calculations of Equations \ref{eq:power_mod} -- \ref{eq:apoiss} agree well with the simulation results for the example shown.\footnote{The slight deficit of simulated power
at high $k$ likely results because the halos are not a perfect Poisson sample (e.g. Smith et al. 2007).} Note that this 
agreement is expected
to the extent that the simulated halo mass function and bias match the respective Sheth-Tormen (2002) and Sheth et al. (2001) 
formulas, and provided that
non-linear biasing and other non-linear effects are negligible or swamped by the Poisson term.

The Poisson term becomes increasingly important as the host halos become more massive
and less abundant. The clustering term also becomes stronger for the more massive halos, but the Poisson term grows more rapidly with mass, and the two terms
become comparable on larger scales (smaller $k$) as the sources become rarer.
For example, in the $M_{\rm co, min} = M_{\rm cool}$ model at $z=7.3$, the Poisson
and clustering terms are comparable at $k$ slightly larger than $1 h$ Mpc$^{-1}$. On the
other hand, if $M_{\rm co, min}$ is as large as $10^{10} M_\odot$, the two terms cross
at $k \approx 0.3 h$ Mpc$^{-1}$ (see Figure \ref{fig:tco_auto}). Note that the models compared in the figure have a fixed duty cycle, $f_{\rm duty} = 0.1$. Decreasing the duty cycle at 
constant $M_{\rm co, min}$ would increase the Poisson term at fixed $\avg{b}^2 P_{\rm lin} (k)$ and Poisson fluctuations would dominate at still larger scales. 
Varying the CO duty cycle also impacts the mean brightness temperature, $\avg{T_{\rm CO}} \propto f_{\rm duty}$, and hence the overall normalization of the 
model power spectra.
The future CO surveys considered in \S
\ref{sec:detectability} potentially probe scales in which each of the clustering and the Poisson
term are important.

\subsection{Parameter Variations}
\label{sec:param_var_auto}

\begin{figure}
\bc
\includegraphics[width=9.2cm]{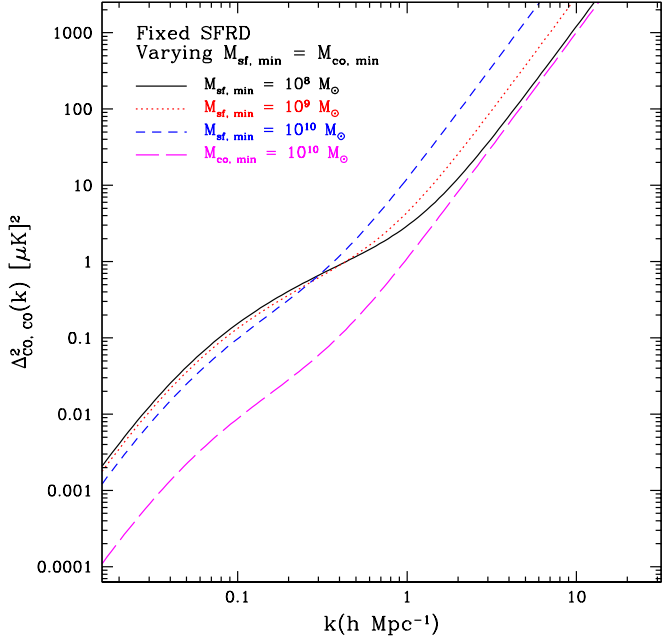}
\caption{The CO auto spectrum for varying $M_{\rm sf, min}$, with $M_{\rm co, min} = M_{\rm sf, min}$ and the SFRD fixed. The SFRD is fixed to 
its value for the model in which the minimum host halo
mass is the atomic cooling mass. The models are at $z=7.3$ and were calculated using the halo model. The magenta dashed line shows a contrasting model where
 star formation occurs in halos down to the atomic cooling mass, but only halos above $M_{\rm co, min} = 10^{10} M_\odot$ are CO luminous.}
 \label{fig:sfrd_fix}
 \ec
 \end{figure}

Before proceeding, let us further examine the impact of some of the uncertainties in our modeling. Given the success of the halo model of Equation
\ref{eq:power_mod} in matching the results of numerical simulations, we will use it in this investigation. The simplest parameter variation to understand
is the impact of uncertainties in the normalization of the $L_{\rm CO}-M$ relation. The average brightness temperature is proportional to this normalization,
and so dialing this value up or down simply results in boosting or diminishing the strength of the brightness temperature fluctuations by the normalization squared.
Next, increasing or decreasing the duty cycle, $f_{\rm duty}$, also boosts/diminishes  
$\avg{T_{\rm CO}}^2$ -- and hence the power spectrum normalization -- as the square of the duty cycle, while
simultaneously varying the level of Poisson fluctuations as $\propto 1/f_{\rm duty}$. For example, for $M_{\rm co, min} = 10^8 M_\odot$
the clustering and Poisson terms are comparable at $k=1 h$ Mpc$^{-1}$ for $f_{\rm duty} = 0.1$, while these terms are comparable
at $k=3 h$ Mpc$^{-1}$ for $f_{\rm duty}=1$.

We have already examined the impact of varying $M_{\rm co, min}$ in Figure \ref{fig:tco_auto}; as with the spatially averaged brightness temperature the
power spectrum is less sensitive to $M_{\rm sf, min}$ itself provided the SFRD is fixed (\S \ref{sec:mean_co}). This is quantified in Figure \ref{fig:sfrd_fix} which
shows the impact of increasing $M_{\rm sf, min}$ above the atomic cooling mass, while fixing both the SFRD and $M_{\rm co, min}=M_{\rm sf, min}$. 
This is accomplished by increasing the normalization of the SFR-M relation (Equation \ref{eq:sfr_mhost}) as $M_{\rm sf, min}$ increases.
Since $\avg{T_{\rm CO}}$ varies with the normalization of the SFR-M relation to the $3/5$th power, this compensates for most of the expected drop in $\avg{T_{\rm CO}}$.
In addition, raising $M_{\rm sf, min}$ increases both the bias factor and the level of Poisson fluctuations which further compensates, and actually leads to the fluctuations
being larger in the high $M_{\rm sf, min}$ models than in the atomic cooling mass model on some scales.

\begin{figure}
\bc
\includegraphics[width=9.2cm]{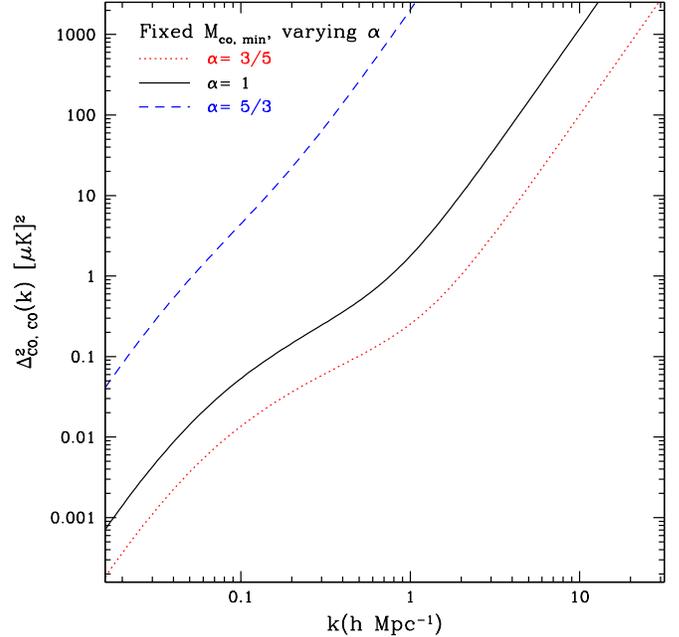}
\caption{The CO auto spectrum for varying $\alpha$. The curves fix $M_{\rm co, min}=10^9 M_\odot$ and $z=7.3$, while varying the power law
index $\alpha$ in the  CO luminosity-halo mass relationship, $L_{\rm CO} \propto M^\alpha$, at constant $L(M_{\rm co, min})$.  Boosting $\alpha$ increases the mean emission,
the clustering strength, and the Poisson fluctuations by giving more weight in the CO emissivity to the rarer, more highly clustered host halos.}
\label{fig:alpha_vary}
\ec
\end{figure}

Finally we vary the power law index in the $L_{\rm CO}-M$ relation, $L_{\rm CO} \propto M^\alpha$. Our fiducial model adopts a linear relationship, $\alpha=1$, which
we choose mainly for simplicity. We also assume that the SFR is a linear function of halo mass, while $z \leq 3$ observations indicate $L_{\rm CO} \propto SFR^{3/5}$ (\S \ref{sec:lum_co}); therefore another sensible model is to assume $L_{\rm CO} \propto M^{3/5}$. Of course, the linear SFR-M relation is also chosen for simplicity -- regardless, the $L_{\rm CO} \propto M^{3/5}$ model illustrates the impact of weighting the low mass halos more heavily and the massive ones less strongly. A contrasting alternative model gives the massive halos more weight, $L_{\rm CO} \propto M^{5/3}$. It is straightforward to generalize the halo model calculation of Equations \ref{eq:power_mod}-\ref{eq:apoiss} for arbitrary power law indices. The results of this calculation are shown in Figure \ref{fig:alpha_vary} for $z=7.3$, the intermediate case of $M_{\rm co, min} = 10^9 M_\odot$, with
$L_{\rm CO} (M=M_{\rm co, min})$ fixed using Equation \ref{eq:lumfid_mhalo}, fixed $f_{\rm duty}$, and each of $\alpha=3/5, 1$, and $5/3$. Increasing $\alpha$ increases the contribution of the massive halos, which 
boosts $\avg{T_{\rm CO}}$, as well as the clustering and Poisson terms. The models with $\alpha=3/5$ and $\alpha=1$ are quite similar, and the impact of decreasing
$\alpha$ is mostly degenerate here with small variations in the $L_{\rm CO}-M$ normalization.  The model with $\alpha = 5/3$, however, produces stronger boosts in the average brightness temperature, as well as the bias and the Poisson term.  In this model, most
of the scales shown in the figure are dominated by the Poisson term. 
The stronger impact of the $\alpha = 5/3$ case can be understood by noting that 
$M^{1+\alpha} dn_{\rm co}/dM \propto M^{-1.2+\alpha}$ near $M = M_{\rm co, min} = 10^9 M_\odot$ at the redshift of interest. Consequently, for $\alpha=3/5, 1$ the logarithmic contribution to the CO emissivity is a monotonically decreasing function of mass above $M_{\rm co, min}$, but this is not the case for $\alpha = 5/3$. At larger minimum mass and/or at higher redshift,
the halo mass function is a more rapidly decreasing function of mass, and increasing $\alpha$ has less impact. On the other hand, the impact of increasing $\alpha$ is
slightly larger when decreasing the minimum mass and/or redshift. In the case that the minimum mass is the atomic cooling mass, the dependence on $\alpha$ is qualitatively similar to the models shown here, but with a more significant boost for $\alpha=5/3$. 

It would be quite interesting to measure CO auto spectra of the type shown in Figures
\ref{fig:tco_auto}-\ref{fig:alpha_vary}. From sufficiently precise measurements one could extract both the
clustering term and the Poisson term, although this will likely be challenging since the
power spectrum is fairly close to a power law in shape.
Nonetheless, it is probably reasonable to assume that the power spectrum is dominated by these terms at large and small scales respectively. 
The clustering term is determined by a product
of the average CO brightness temperature and the bias factor of the host halos, while
the Poisson term depends on the average brightness temperature, and the abundance/duty
cycle of CO luminous galaxies.
This would provide valuable information on which
types of galaxies are CO luminous at high redshift, how massive their host halos are,
and what the duty cycle of CO-luminous activity is in these halos.
Particularly notable is that this
measurement would constrain the cumulative emission from many very faint, unresolved
galaxies, which individually may be too faint to detect even with deep observations from ALMA. 
More generally, the measurement would trace the properties of molecular clouds -- i.e., the locations
where stars form -- 
in the very galaxies that reionized the Universe.

It would also be interesting to investigate the angular dependence of the CO auto power spectrum. The
power spectrum will be anisotropic owing to redshift space distortions from peculiar velocities 
(Kaiser 1987), and measurements of this anisotropy can be used to break model degeneracies.
A sufficiently precise measurement of the quadrupole to monopole ratio of the power spectrum, for example,
determines the quantity $\beta = \Omega_m(z)^{0.6}/b \approx 1/b$ for the CO emitting galaxies. In principle, combining this measurement
with the spherically averaged power spectrum, allows one to {\em separately} determine each of $\avg{T_{\rm CO}}$ and $b$
from low $k$ measurements, and then infer the Poisson term at high $k$.

Yet another interesting quantity to measure is the ratio of the power spectrum of fluctuations in the
CO(2-1) and CO(1-0) lines. The ratio of the root-mean-squared fluctuations in the two lines gives a (fluctuation-weighted)
measure of the excitation temperature and probes the heating rate in the reionizing sources.

\section{Cross correlation with Redshifted 21 cm Emission}
\label{sec:cross_co}

Perhaps the most exciting prospect for CO intensity mapping, however, is to combine
it with future observations of the redshifted 21 cm line from the high
redshift IGM. As motivated in the introduction, the cross spectrum should be less
sensitive to systematic effects and provide complementary information about the EoR to the
21 cm auto spectrum.

For simplicity, we assume that the spin temperature of the 21 cm transition is much larger
than the CMB temperature globally, which is likely a good approximation during most of the 
EoR (Ciardi \& Madau 2003, Pritchard \& Furlanetto 2007). Further, we ignore the impact of peculiar
velocities which should not significantly influence our present calculations (Mesinger \& Furlanetto 2007).
With these assumptions, the 21 cm brightness temperature at spatial position $\r$ can be written as (e.g. Zaldarriaga et al. 2004):
\beqa
T_{\rm 21}(\r) = 28 mK \avg{x_{HI}} \left[1 + \delta_x(\r)\left]\right[1 + \delta_\rho(\r)\right] \left[\frac{1+z}{10}\right]^{1/2}. \nonumber \\
\label{eq:t21}
\eeqa
Here $\avg{x_{HI}}$ denotes the volume-averaged neutral hydrogen fraction, $\delta_x(\r)$ denotes the fractional fluctuation in neutral hydrogen
density at spatial position $\r$, and $\delta_\rho(\r)$ is the fractional gas density fluctuation. We will also use the 
symbol $\avg{x_i} = 1 - \avg{x_{\rm HI}}$ to
denote the volume-averaged ionization fraction. The timing and duration of reionization are still quite uncertain, and so the redshift at which a given fraction
of the IGM volume is ionized may be different than in our particular reionization model. However, Furlanetto et al. (2006b) and McQuinn et al. (2007a) show 
that the size
of the ionized regions during reionization depend mostly on the ionized fraction, $\avg{x_i}$, rather than the precise redshift at which a given volume is 
ionized.
As a result, the shape of the cross spectrum at a given ionization fraction is likely
a more robust prediction than that at a given redshift.

\begin{figure}
\bc
\includegraphics[width=9.2cm]{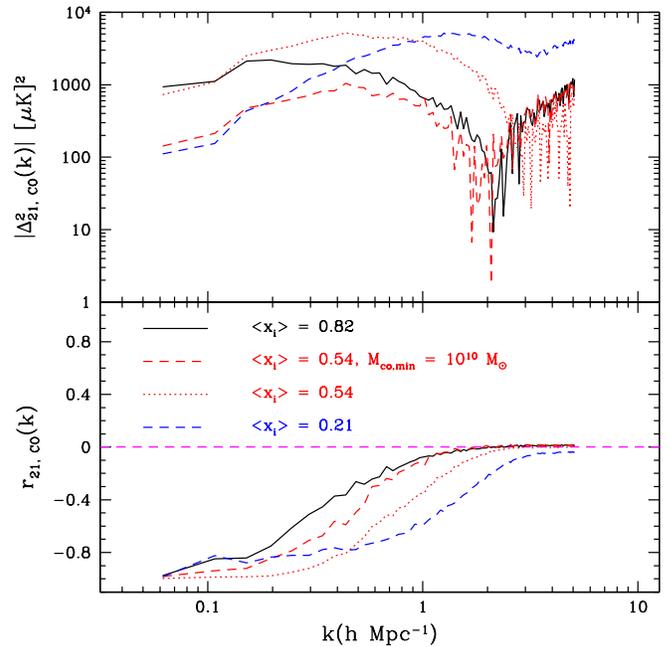}
\caption{The cross power spectrum between the CO and 21 cm brightness temperature
fluctuations. {\em Top panel}: The absolute value of the cross spectrum between
21 cm and CO emission in units of $(\mu K)^2$ at different redshifts and ionization 
fractions. The redshifts at the corresponding ionization fractions are $(z, \avg{x_i}) = 
(6.90,0.82);(7.32,0.54);(8.34,0.21)$. The red dashed line adopts $M_{\rm co, min} = 10^{10}
M_\odot$, while the other curves assume that halos down to the atomic cooling mass host
CO-luminous sources. {\em Bottom panel}: The cross-correlation coefficient between the two
fields as a function of wavenumber.}
\label{fig:power_cross}
\ec
\end{figure}

Using Equation \ref{eq:t21}, we produce three-dimensional maps of the 21 cm field from outputs of the reionization simulations at various redshifts/ionized fractions.
We then measure the cross power spectrum between the 21 cm and CO data cubes as described in the previous section.
The results of these calculations are shown in Figure \ref{fig:power_cross}. The top panel shows the absolute value
of the cross spectrum, while the bottom panel indicates the cross correlation coefficient between the two
random fields as a function of wavenumber, $r_{\rm 21, CO}(k)$. The cross correlation coefficient is defined by 
$r_{\rm 21, CO}(k) = P_{\rm 21, CO}(k)/\left[P_{\rm CO, CO}(k) P_{\rm 21, 21}(k)\right]^{1/2}$ and is $1$ ($-1$) for wavenumbers
in which the two fields are perfectly correlated (anti-correlated), while it is zero for wavenumbers in which the two fields
are completely uncorrelated.

The simulated cross spectra are similar to those in Lidz et al. (2009), and we refer the reader to this paper for a more detailed discussion, but
summarize some of the main features here.\footnote{The main difference with Lidz et al. (2009) is that the 
CO emission in our model is proportional to the mass-weighted halo abundance, while these authors' galaxy density field is directly
proportional to the halo abundance.} 
On large scales, the 21 cm and CO fields are anti-correlated. To understand this, consider length scales larger than the size
of the ionized bubbles during a given stage of reionization. Regions that are overdense on large scales contain more galaxies, and are hence
brighter in CO emission than typical regions. The same regions, however, correspond to mostly ionized portions of the 21 cm map, and
are consequently dimmer than average in 21 cm emission.
On these spatial scales,
the two fields are consequently anti-correlated. On the other hand, the two fields are {\em uncorrelated} on scales smaller than the
ionized bubbles around groups of CO-emitting galaxies. This occurs because the gas at each point within an ionized region is highly ionized
irrespective of the precise galaxy density. Similarly, fully neutral regions do not contain galaxies (unless
some galaxies have very low ionizing photon escape fractions). 

The cross-correlation coefficient, $r_{\rm 21, CO}(k)$, in Figure \ref{fig:power_cross} illustrates exactly these trends. On large scales,  the simulated $r_{\rm 21, CO}(k)$ goes
to $r_{\rm 21, CO}(k) = -1$, while it drops to zero on small scales. The scale where $r_{\rm 21, CO}(k)$ goes to zero increases with decreasing redshift as the Universe
becomes progressively more ionized and the ionized regions grow. The red dashed line shows that this behavior is sensitive, however, to
the minimum mass of CO-luminous galaxies. In the red-dashed line model $M_{\rm co, min}$ is larger than the fiducial value, ($M_{\rm co, min} = 10^{10} M_\odot$ rather than
$M_{\rm co, min} = M_{\rm cool}$),  and the cross spectrum turns over on larger scales. This happens because the more massive halos are more clustered,
and hence tend to be surrounded by larger bubbles than the less massive halos. Although the ionized regions at the stages of reionization considered here are much
larger than the size of HII regions around individual galaxies, it is still the case that more massive halos live in larger overdensities and tend to be surrounded
by larger ionized regions. In order to interpret the cross spectrum's turnover scale unambiguously, one hence needs to separately constrain $M_{\rm co, min}$, which
may be possible with measurements of the CO auto spectrum.

In principle, the turnover in the cross spectrum between a `traditional' galaxy survey, with resolved galaxies, and the redshifted 21 cm signal, may be easier
to interpret than the intensity mapping signal considered here. However, as discussed in the introduction, traditional galaxy surveys are poorly matched to the
redshifted 21 cm observations. Furthermore, surveys for high redshift LAEs are the only high redshift galaxy surveys anticipated in the near future 
with sufficiently wide fields of view.
The observed abundance of LAEs will be modulated by surrounding neutral hydrogen in the IGM, which will in turn impact the turnover in the cross
spectrum (Lidz et al. 2009), complicating the interpretation of this measurement. Hence the 21 cm-CO cross spectrum may indeed be the most promising approach. 
 
Let us now consider the amplitude of the cross power spectrum, as shown in the top panel of Figure \ref{fig:power_cross}. The amplitude of the cross spectrum at 
$k \sim 0.1 \hmpc$ varies from about $10^2 (\mu K)^2$ to about $10^3 (\mu K)^2$, depending on redshift and model.  These amplitude variations reflect several
different aspects of the signal, the most important of which are that the cross spectrum is proportional to $\avg{T_{\rm CO}}$, and that the 21 cm signal depends strongly
on $\avg{x_i}$. The dependence on $\avg{T_{\rm CO}}$ means that the CO emission falls off towards high redshift, and with increasing $M_{\rm co, min}$, although 
these effects are partly compensated by increases in $\avg{b}$ with host halo mass and redshift. This behavior is similar to that of 
the CO auto spectrum (\S \ref{sec:co_auto}), except that the cross spectrum is linear in $\avg{T_{\rm CO}}$
and $\avg{b}$, while the auto spectrum is quadratic in these quantities. The 21 cm fluctuations are generally largest in amplitude on most relevant scales around reionization's mid-point,
when $\avg{x_i} = 0.5$ (e.g. Lidz et al. 2008). This results because the ionized bubbles grow larger as reionization proceeds, which initially boosts the large scale 21 cm power, while
the 21 cm fluctuations must eventually fall off as the Universe becomes completely ionized. In the particular reionization model shown, $\avg{x_i} \approx 0.5$ at $z \approx 7.5$, and the
cross spectrum peaks at this redshift for most relevant wavenumbers. If reionization's mid-point occurs at a higher redshift than in this model, the peak cross spectrum amplitude
may be reached at a later stage of reionization -- i.e., at a larger $\avg{x_i}$, $\avg{x_i} \gtrsim 0.5$. In particular, if reionization's mid-point occurs at a sufficiently high redshift, 
the boost in 21 cm power towards $\avg{x_i} \sim 0.5$ may be overcome by the fall of in $\avg{T_{\rm CO}}$ with increasing redshift. This would make it more challenging
to detect the cross spectrum signal.
However, if reionization occurs
earlier than in our model, the SFRD would also be larger at high redshift and so $\avg{T_{\rm CO}}$ may not fall off so rapidly. At any rate, this discussion illustrates
that measuring the cross spectrum amplitude as a function of redshift provides additional information, and is another example
of how the cross spectrum measurement can complement 21 cm auto spectrum measurements.

We now have a sense for the amplitude and shape of the 21 cm-CO cross spectrum, and its dependence on model, ionized fraction, and redshift.
If it is practical to measure this signal, it can be used to determine the size of the ionized regions around CO-emitting galaxies. If the measurement is feasible over
a range of redshifts and hence at various stages of the reionization process, it would provide a direct tracer of the growth of ionized regions during the EoR.
Combining this with measurements of the CO auto spectrum can potentially break degeneracies owing to uncertainties in the minimum
host halo mass of CO emitting galaxies, and other uncertainties in the properties of the CO emitting galaxies.

\section{Foreground emission and combining different CO transitions}
\label{sec:trans_co}

Let us move to consider the impact of foreground contamination, which is one important factor in assessing the practicality of measuring the CO auto spectrum and the 21 cm-CO cross spectrum. 
Cleaning the CO foregrounds should proceed essentially as planned for the redshifted 21 cm signal: one can distinguish foregrounds from the underlying
signal by using the fact that the foregrounds are spectrally smooth, while the signal has structure in frequency space (e.g. Zaldarriaga et al. 2004, Morales \& Hewitt 2004,
McQuinn et al. 2006, Petrovic \& Oh 2010). This will separate the CO emission line signal from the much larger `foreground' emission from
our own galaxy, the primary CMB, point source emission, and other sources. In order to give a rough idea of the fidelity at which foregrounds must be removed in the CO data
relative to the redshifted 21 cm case, it is useful to consider the ratio of the mean foreground emission to the mean signal for each of the 21 cm and CO signals/observing
frequencies. Adopting $1 \mu K$ for the mean CO emission signal and $200 \mu K$ for the sky temperature at $\sim 30$ Ghz (e.g. de Oliveira-Costa et al. 2008), the ratio of foreground
to mean signal is $\sim 200$ for the CO observations, compared to $\sim 10^4$ for the redshifted 21 cm observations. From this standpoint, cleaning foreground emission from the CO observations 
appears less demanding than it does for the redshifted 21 cm signal. 
 
Note that there is, however, one slight disadvantage of cross correlating 21 cm and CO maps as compared to the possibility of cross-correlating 21 cm and optically-selected
galaxies, such as LAEs. The disadvantage is that
CO measurements at $\sim 15-30$ Ghz will {\em share} much of the same foreground structure that impacts the 21 cm data cubes, unlike the case of the LAEs. For example,
de Oliveira-Costa et al. (2008) find that a principle component model with only three components can fit maps of diffuse Galactic radio emission at frequencies
from 10 Mhz to 100 Ghz, suggesting a highly-correlated origin for Galactic foreground emission at these frequencies.

However, the correlated nature of the CO and 21 cm foregrounds is unlikely a big concern. First, cleaning the spectrally smooth component of each of the CO and 21 cm data should
remove the contamination extremely efficiently.  
For instance, for the 21cm case, both analytic (McQuinn et al 2006) and simulation (Petrovic \& Oh 2011) calculations show that the foreground residuals after cleaning are minuscule, and negligible compared to the signal. The main systematic in fact arises from excess cleaning, which results in reduction of large scale power. 
A conservative check to eliminate the possibility of correlated foregrounds can be performed simply by cranking up the strength of the foreground cleaning, e.g. by fitting the foregrounds with a progressively higher
order polynomial. Moreover, if one can make three-dimensional intensity maps in each of the CO(2-1) and the CO(1-0) lines for gas emitting at a given redshift, this can provide an additional cross-check.
For instance, one could take a wavelength-weighted difference of maps in the CO(2-1) and CO(1-0) lines to remove the highly-correlated foregrounds in these maps, and
correlate this difference map with the redshifted 21 cm signal. 

Using two rotational lines from gas emitting at a given redshift can also protect against 
the possibility
of foreground `interlopers' -- i.e., line emission from lower redshift galaxies at the observed frequencies of interest. The strongest interloper line for the CO(2-1) transition is likely to be CO(1-0) emission from
gas at lower redshift. Note that contamination from high redshift CO(2-1) emission for CO(1-0) intensity mapping during the EoR should be negligible.
Other possible foreground interlopers are emission from HCN molecules and radio recombination lines, but these are faint 
compared to the CO emission.
Regardless, cross correlating data cubes in the CO(1-0) and CO(2-1) lines with each other and with the redshifted 21 cm data cubes should protect against the 
interloper lines. Visbal \& Loeb (2010) consider multiple emission line (fine structure lines, CO lines, and 21 cm) tracers of large scale structure
after the EoR and consider using cross correlations to evade interloper contamination in more detail. It would be interesting
to investigate this further, especially given the uncertainties in the ratio of the luminosity in the CO(1-0) and CO(2-1)
lines, and the potential scatter in this relationship.

\section{Detectability}
\label{sec:detectability}

Now that we have theoretical predictions for the CO auto spectrum and the 21 cm-CO cross spectrum and have discussed systematic 
effects from foreground contamination, we turn to consider the detectability of these signals.
Let us start by considering the CO auto spectrum. First, we consider the CO noise power spectrum rather generically, and then we 
discuss the type of instrument that will be needed to detect the signal more concretely (\S \ref{sec:instrument}).

\subsection{Detectability of the CO auto spectrum}
\label{sec:auto_detect}

We assume that the CO thermal noise power spectrum is a pure white-noise spectrum. The CO brightness temperature 
fluctuations are typically of order $\sim 1 (\mu K)^2$, although
with significant dependence on model, redshift, and spatial scale. Owing to this, we characterize the CO noise by the noise variance in
$10$ arcminute spatial pixels and spectral channels of width $\Delta \nu/\nu = 1\%$; we denote this variance by $\sigma_N$ and choose values
around $\sigma_N \sim 1 \mu K$. The size of the spatial pixels and the spectral channels are just chosen as convenient numbers at which to quote 
the noise variance;
in practice we find that slightly smaller pixels are preferable.  In the case of a pure white-noise random field the noise power spectrum may be written as
\beqa
P_{\rm N,CO} = \sigma_N^2 V_{\rm pix},
\label{eq:var_co}
\eeqa
with $V_{\rm pix}$ denoting the co-moving volume corresponding to $1 \%$ spectral channels that are $10$ arcminutes on a side. For reference, this corresponds
to $6.75 \times 10^3$ (Mpc/$h$)$^3$ at $z=7$.

Assuming Gaussian statistics, the variance of the CO auto spectrum for a single $k$-mode, with line of sight component $k_\parallel = \mu k$ and transverse component $k_\perp^2 = k^2 - k_\parallel^2$,
is:
\beqa
\rm{var}\left[P_{\rm CO}(k,\mu)\right] = \big[P_{\rm CO}(k) + \nonumber \\
P_{\rm N,CO}(k) e^{(k_\parallel/k_{\parallel, res})^2  + (k_\perp/k_{\perp, res})^2} \big]^2. 
\label{eq:var_co}
\eeqa
The first term on the right hand side of this equation is the usual sample variance term, while the second one comes from thermal noise in the telescope.
The exponential reflects the limited spectral and spatial resolution of the instrument. Here $k_{\parallel,res}$ denotes smoothing from finite spectral resolution, and
$k_{\parallel,res} = H(z)/[c (1+z)] (\nu_{\rm obs}/\Delta \nu_{\rm obs})$; while $k_{\perp, res} = 2 \pi/(D(z) \theta_{\rm min})$ is the spatial smoothing with
$D(z)$ denoting the co-moving distance to the redshift of interest, and $\theta_{\rm min}$ giving the angular size of the spatial pixels.

In addition, we would like to calculate the variance of the spherically-averaged power spectrum.
We consider logarithmic bins of width
$\epsilon = \rm{d ln} k$. In this case, the minimum variance estimate of the spherically averaged power spectrum
has a variance of
\beqa
\frac{1}{\sigma^2_P(k)} = \sum_\mu \frac{k^3 V_{\rm survey}}{4 \pi^2} \frac{\Delta \mu}{\sigma^2_P(k,\mu)}.
\label{eq:sigma_p} 
\eeqa
The sum here runs over the upper half-plane (i.e., positive $\mu$) because we consider the power spectrum of a real-valued
field, and only half of the Fourier modes are independent as a result. It is helpful to note that in the case that the noise and signal power
spectra are $\mu$-independent, the above formula simplifies to
$\sigma_P^2(k) = [P(k) + P_N(k)]^2/N_m$, where $N_m$ is the number of modes in a $k$-bin (counting only modes in the upper half plane).
Here $V_{\rm survey}$ denotes the co-moving volume of the CO survey. If the survey samples a depth $\Delta D$, centered
on a redshift $z$, and covers a field of view on the sky of $\Omega_S$ steradians, then 
$V_{\rm survey} = \Omega_S D(z)^2 \Delta D$. The sum over $\mu$ is restricted by the survey dimensions.

In order to best measure the 21 cm-CO cross spectrum, the CO and 21 cm surveys should aim to have similar $k$-mode coverage.
We anticipate that this requirement will fuel the design of the telescopes planned for the CO measurement.
To achieve this, the CO survey will want a wide field of view and several arcminute spatial resolution. For our baseline numbers, we assume that the
CO survey covers $25$ deg$^2$ on the sky, and that $\Delta D = 68.6$ Mpc/$h$, corresponding to a bandwidth of $6$ Mhz for the 21 cm survey
we consider shortly. This field of view is comparable to that of LOFAR (Harker et al. 2010), but less than the $\sim 800$ deg$^2$ planned for the MWA.
We assume that each spatial pixel has a size of $\theta_{\rm min} = 6$ arcminutes, corresponding to $k_{\perp, res} = 0.58 h$ Mpc$^{-1}$ at $z=7$.
For reference, the entire CO survey spans a co-moving volume of
$2.1 \times 10^7$ (Mpc/$h$)$^3$ co-moving at $z=7$. 

\begin{figure}
\bc
\includegraphics[width=9.2cm]{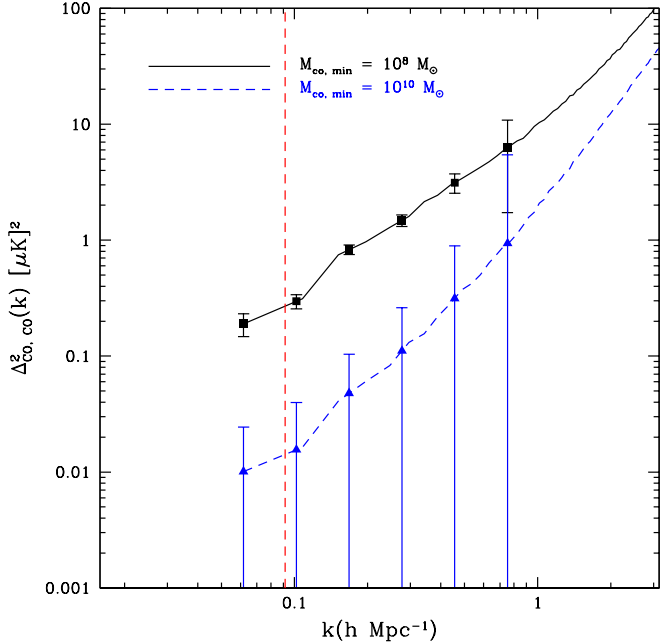}
\caption{Error bar estimates for the CO power spectrum at $z=7$. The black squares show error bar estimates for the spherically averaged
CO power spectrum at $z=7$ assuming a theoretical model with $M_{\rm co, min} = 10^8 M_\odot$, while the blue triangles are for
$M_{\rm co, min} = 10^{10} M_\odot$. The CO survey covers a field of view of $25$ deg$^2$,
with $\sigma_N = 1 \mu K$, spectral channels of  width $\Delta \nu = 0.05$ Ghz, and spans a depth of $68.6$ co-moving Mpc/$h$ (see text).  Scales roughly to the
left of the red dashed line will be lost to foreground cleaning, while small scales are lost owing to the limited spatial and spectral resolution of the instrument.}
\label{fig:errb_co_auto}
\ec
\end{figure}

Figure \ref{fig:errb_co_auto} shows error bar estimates for two simulated models at $z=7$. The simulated models have $M_{\rm co, min} = 10^8 M_\odot$ and 
$10^{10} M_\odot$ respectively.
The power spectrum has been averaged in spherical bins of width $\epsilon = 0.5$. As mentioned above,
the CO intensity mapping experiment is assumed to cover $25$ deg$^2$ on the sky, with a thermal noise of $\sigma_N = 1 \mu K$ (quoted
as the equivalent noise in $10$ arc-minute pixels, and $1\%$ spectral channels), in spectral pixels of width $\Delta \nu = 0.05$ Ghz, or
$\Delta \nu/\nu = 1.7 \times 10^{-3}$, and $6$ arcminute spatial resolution.\footnote{For reference, note that if the pixel noise is $\sigma_N = 1 \mu K$ for $10$ arc-minute spatial pixels, 
and $1\%$ spectral channels,
the corresponding noise variance in $6$ arcminutes, $\Delta \nu = 0.05$ Ghz pixels is $\tilde{\sigma}_N = 4 \mu K$. This is obtained assuming pure white noise and using Equation \ref{eq:var_co}.}
 The results of the $M_{\rm co, min} = 10^8 M_\odot$ model look encouraging, provided
these survey parameters are achievable: high
significance CO power spectrum measurements are possible across roughly a decade in spatial scale, $k \sim 0.1-1 h$ Mpc$^{-1}$. The fractional error bars
are larger in the $M_{\rm co, min} = 10^{10}$ model, which has smaller amplitude fluctuations. This quantifies the difficulty in detecting CO fluctuations
in scenarios in which galaxies with low SFRs are dim in CO. If this model is representative, a CO survey with smaller thermal noise is required for a firm
measurement. Note again the distinction between $M_{\rm sf, min}$ and $M_{\rm co, min}$ in our modeling: if all star-forming galaxies are CO luminous --
i.e., $M_{\rm co, min} = M_{\rm sf, min}$ --
and the SFRD is fixed, then there is relatively little sensitivity to increasing $M_{\rm sf, min}$ itself (\S \ref{sec:param_var_auto}).  
If the CO luminosity-halo mass normalization (Equation \ref{eq:lumfid_mhalo}) is a factor of $\sim 5$ lower than in our fiducial model, this
would result in a similar reduction in $S/N$ as results from increasing $M_{\rm co, min}$ from $10^8 M_\odot$ to $10^{10} M_\odot$. 

In the $M_{\rm co, min} = 10^8 M_\odot$ model, the finite spatial and spectral resolution prevent measurements on small spatial scale, while
foreground contamination will inevitably limit ones ability to measure CO fluctuations on large spatial scales. As discussed in \S \ref{sec:trans_co}, one will want
to remove a spectrally smooth component from the CO data to separate the CO emission line signal from the much larger spectrally smooth foregrounds.
The precise impact of the foreground cleaning process will depend on the bandwidth over
which the spectrally smooth component is fit, the precise cleaning algorithm, and the spectral shape of the foregrounds. As a rough guide, it is useful to note that if the cleaning is done over a bandwidth corresponding to a co-moving depth of $\Delta D$,  {\em at least} all modes
with $k \leq 2 \pi/\Delta D$ will be strongly impacted by foreground cleaning. In fact, modes on smaller scales will have suppressed power due to aliasing effects, though if correctly handled one can still obtain unbiased estimates of the power spectrum at these scales, albeit with 
somewhat larger variance (Petrovic \& Oh 2011, Liu \& Tegmark 2011). For example, future 21 cm surveys will likely measure power spectra over a bandwidth 
of $B \sim 6$ Mhz
to avoid evolution in the signal across the observed bandwidth (e.g. McQuinn et al. 2006). 
If the foregrounds are cleaned over this same bandwidth, this will strongly impact measurements
for modes of length scale larger than $\Delta D = 68.6$ Mpc/$h$, or $k \leq 0.092 h$ Mpc$^{-1}$. This scale is indicated by the red dashed line in the figure.  One might use a larger bandwidth in cleaning the foregrounds, but in this case one will
generally require a larger number of parameters to adequately describe the foregrounds. This will ultimately limit the gain from using a larger bandwidth in the cleaning algorithm (McQuinn et al. 2006).
Since the CO power spectrum should evolve less with redshift than the 21 cm power spectrum during reionization, one can likely extract the CO auto spectrum on
somewhat larger scales than the 21 cm power spectrum. Nonetheless, the dashed red line should provide a conservative estimate of the scales that are strongly 
affected by
foreground cleaning. 
Measuring the CO auto spectrum over even this decade in scale would provide interesting constraints on $\avg{T_{\rm CO}}$, $\avg{b}$ and the Poisson term, which would in turn
constrain the star formation rate density at high redshift, and the properties of CO emitting gas at very early cosmic times.

\subsection{Detectability of the 21 cm-CO cross spectrum}
\label{sec:cross_detect}

\begin{figure}
\bc
\includegraphics[width=9.2cm]{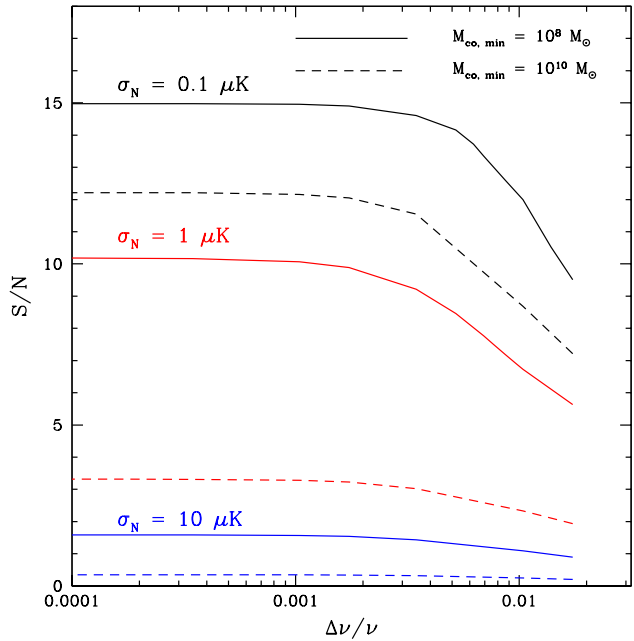}
\caption{Detectability of 21 cm-CO cross spectrum at $z=7$, $\avg{x_i} = 0.82$ as a function of spectral resolution and pixel noise level for the CO measurement. The curves show the total
$S/N$ (summed over all $k$-modes) at which the 21 cm-CO cross spectrum may be detected. An $S/N$ of $5$ indicates, for example, that the cross spectrum can be
detected at $5-\sigma$ confidence. The solid lines are for $M_{\rm co, min} = 10^8 M_\odot$, while the dashed lines take $M_{\rm co, min} = 10^{10} M_\odot$.}
\label{fig:ston_21cm_co}
\ec
\end{figure}
We now consider the detectability of the 21 cm-CO cross spectrum, which is potentially an even more interesting signal to detect.  The variance of the cross-spectrum estimate
for a single Fourier-mode is a generalization of Equation \ref{eq:var_co} (e.g. Furlanetto \& Lidz 2007):
\beqa
\rm{var}\left[P_{\rm 21, CO}(k,\mu)\right] = \frac{1}{2} \left[P_{\rm 21, CO}(k,\mu) + \sigma_{\rm CO}(k,\mu) \sigma_{\rm 21}(k,\mu)\right]^2. \nonumber \\
\label{eq:var_cross}
\eeqa
In this Equation, $\sigma_{\rm CO}(k,\mu)$ and $\sigma_{\rm 21}(k,\mu)$ are short-hands for the error bar on the CO and 21 cm auto spectra respectively for a mode with
wavenumber $k$ and $\mu = k_\parallel/k$, while
$P_{\rm 21,CO}(k,\mu)$ denotes the cross power spectrum. In order to evaluate this, we need to first compute the variance of the 21 cm auto spectrum which is
given by (McQuinn et al. 2006):
\beqa
\rm{var}\left[P_{\rm 21}(k,\mu)\right] = \left[P_{\rm 21}(k,\mu) + \frac{T^2_{\rm sys}}{B t_{\rm int}} \frac{D^2 \Delta D}{n(k_\perp)} \left(\frac{\lambda^2}{A_e}\right)^2\right]^2.\nonumber \\
\label{eq:var_21}
\eeqa
The second term in this expression is the 21 cm noise power spectrum, while the first term is the usual sample variance term. In this equation, $T_{\rm sys}$ is the
system temperature, $B$ is the bandwidth, $t_{\rm int}$ is the integration time, $\lambda$ is the observed wavelength, $A_e$ is the effective area of each antenna tile, 
and $n(k_\perp)$ denotes the number density of baselines observing a mode with
transverse wavenumber $k_\perp$ (McQuinn et al. 2006, Bowman et al. 2006, Lidz et al. 2007). In order to estimate the error bar on the spherically averaged
cross spectrum we sum over modes in logarithmic $k$-bins, as in Equation \ref{eq:sigma_p}, except restricting the survey volume and the sum to modes that are common to both the CO and 21 cm surveys.

We consider the prospects for cross-correlating three-dimensional CO maps with 21 cm measurements from the MWA. Similar considerations and calculations could be performed for the other first generation 21 cm arrays. The MWA will consist of $500$ antenna tiles each with an effective
area of $A_e = 14 m^2$ at $z=8$ (Bowman et al. 2006). We assume that a fraction of the antennas are packed as closely as possible within a $20$ m core, and that
the remaining antennas follow an $r^{-2}$ distribution out to a maximum baseline of $1.5$ km (Bowman et al. 2006). We take $B=6$ Mhz and $t_{\rm int} = 1,000$ hrs,
and $T_{\rm sys} = 325$ K. The MWA will survey a large field of view on the sky of $\sim 800$ deg$^2$; we assume that only the $25$ deg$^2$ patch of this covered by the CO
survey is available for estimating the 21 cm-CO cross spectrum.

Figure \ref{fig:ston_21cm_co} shows an estimate of the total $S/N$, summed over all $k$-modes, at which the 21 cm-CO cross power spectrum can be detected for various
CO survey parameters. These estimates are shown for models with each of $M_{\rm co, min} = 10^8 M_\odot$ and $10^{10} M_\odot$. The x-axis shows the spectral resolution
of the CO measurement, $\Delta \nu/\nu$. For $\sigma_N = 0.1 \mu K$, significant ($S/N \gtrsim 10-\sigma$ confidence) cross spectrum detections appear feasible in both models. If $\sigma_N = 1 \mu K$ then a significant detection is only feasible in the atomic cooling mass model, while if $\sigma_N = 10 \mu K$, neither model
yields a high $S/N$ measurement. Increasing the spectral resolution is initially helpful, but the gains saturate around $\Delta \nu/\nu = 0.003$ corresponding to
$k_{\parallel, res} \sim 1 h$ Mpc$^{-1}$ since the finite spatial resolution and thermal noise prohibit measuring higher $k$-modes. 
Note that the dependence on spectral resolution is not so strong; if the
total number of spectral channels that is observationally feasible is fixed, it may be advantageous to increase the observational bandwidth
at fixed spectral resolution.

\subsection{CO Instrumental Configuration}
\label{sec:instrument}

From the above estimates, it appears that significant CO auto spectrum detections and 21 cm-CO cross spectrum detections are possible if suitable CO surveys
are feasible. In the case that a significant fraction of the ionizing sources are CO dim (e.g., $M_{\rm co, min} \gtrsim 10^{10} M_\odot$), a detection will be
more challenging, but interesting upper limits might be placed. These statements hinge on the feasibility of a CO survey covering
$\sim 25$ deg$^2$ on the sky with $\sim 6$ arcminute, $\Delta \nu/\nu \sim 0.003$ pixels, at a thermal noise level of $\sigma_N = 0.1-1 \mu K$ (the noise level is quoted for coarser $10$ arcminute, $1 \%$ spectral pixels).

Here we briefly consider what sort of interferometer specifications are needed to match these requirements. 
In order to map $25$ deg$^2$ on the sky in the
CO(2-1) line at z=7, $\lambda_{\rm obs} = 1$ cm, antennas of size $D \sim 12$ cm are required. In order to achieve spatial resolution of $\sim 6$ arcminutes,
maximum baselines of length $D_{\rm max} \sim 6$ m are necessary. Assuming for starters uniform coverage in Fourier space, the thermal noise in an image plane pixel is
\beqa
\sigma_N = \frac{T_{\rm sys}}{\sqrt{\Delta \nu t_{\rm int}}} \frac{1}{f_{\rm cover}}.
\label{eq:var_21}
\eeqa
In this equation $f_{\rm cover}$ denotes the covering factor of the $N_a$ antennas, $N_a D^2/D^2_{\rm max}$. We can then estimate the covering factor required
to achieve $\sigma_N \sim 1 \mu K$ in $10$ arcminute, $1 \%$ spectral pixels. Assuming $T_{\rm sys} = 30$ K and $t_{\rm int} = 1,000$ hrs requires $f_{\rm cover} \sim 1$
to reach $\sigma_N = 1 \mu K$.
For strictly uniform Fourier coverage this amounts to $N_a \sim 900$ antennas, but this is surely a very conservative estimate; in reality the antennas will not
be uniformly distributed and this is likely to boost the sensitivity significantly. Furthermore, the baselines are small enough to rotate the entire telescope and help fill
in the coverage in Fourier space.
More detailed investigation is clearly warranted, but it seems likely that a few hundred antenna elements would suffice for the CO measurement. An alternative observational
strategy is to build a focal plane array rather than an interferometer (Bowman et al. 2011, in prep.); it would be interesting to  compare these approaches in detail. 

\subsection{21 cm-CO correlations using next generation 21 cm Surveys}

Since high redshift CO intensity mapping is a relatively new idea (starting with Righi et al. 2008),
CO intensity mapping instruments may be built on the timescale of second generation
21 cm survey instruments. Here we briefly consider the prospects for measuring
the 21 cm-CO cross spectrum using second generation 21 cm surveys. In order
to optimize the $S/N$ of this measurement, the two surveys should strive
to have similar coverage of Fourier modes. These considerations should shape the
design of each instrument. To name one example, if the CO surveys are necessarily limited
to fields of view that are tens of square degrees, then a 21 cm array design like LOFAR
which has a smaller field of view than the MWA but more collecting area, is advantageous.

\begin{figure}
\bc
\includegraphics[width=9.2cm]{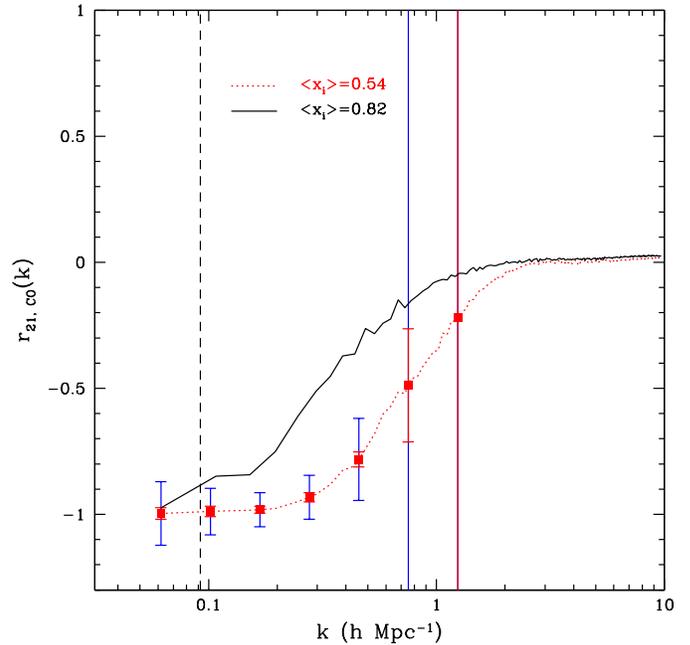}
\caption{Cross correlation coefficient between the 21 cm and CO fields with error bars for a futuristic 21 cm survey. The curves
and error bars are at $z=7$. The red dotted line and associated 1-$\sigma$ error bars assume $\avg{x_i} = 0.54$, while the black solid line shows a
contrasting model with $\avg{x_i} = 0.82$. The red error bars assume that the thermal noise in the CO survey is $\sigma_N = 0.1 \mu K$, while the larger blue
error bars take $\sigma_N = 1 \mu K$. 
The two
models can be distinguished at high significance. Scales roughly to the left of the dashed line are strongly affected by foreground
contamination, while the error bars blow up on small scales owing to the limited spatial and spectral resolution of the CO survey.} 
\label{fig:rco_mwa5000}
\ec
\end{figure}

To illustrate the plausible sensitivity of these future measurements, we assume a CO survey
covering $25$ deg$^2$, with $\Delta \nu = 0.05$ Ghz and consider each of $\sigma_N = 0.1 \mu K$ and 
$\sigma_N = 1 \mu K$. For the
21 cm survey, we consider an array similar to the MWA, except with $5,000$ total antennas. The antennas are initially 
arranged as close together as possible
over an $80$ m core, followed by the usual $r^{-2}$ distribution out to maximum baselines
of $1.5$ km. This is identical to the futuristic version of the MWA dubbed `MWA 5000' by McQuinn et al. (2006). 
We consider the ability of these instruments to detect the turnover in the cross-correlation coefficient at $z=7$. Here
we adopt a fiducial model with $M_{\rm sf, min} = 10^8 M_\odot$ and $\avg{x_i}=0.54$. The cross correlation coefficient for this
model, with error bars is shown by the red dotted line with error bars in Figure \ref{fig:rco_mwa5000}. The black solid line
is a contrasting model where reionization has progressed further by $z=7$ (to $\avg{x_i} = 0.82$) and the ionized regions are larger,
causing the cross correlation coefficient to turn over on larger scales in this model.
These two models can clearly be distinguished at high significance for either level of thermal noise in the CO survey, further illustrating the potential promise of the
21 cm-CO cross correlation measurements. 
The error bars increase rapidly towards small scales because of the limited
spatial and spectral resolution of the assumed CO survey, and so improvements here could allow even better measurements.

\section{Comparison with Previous Work}
\label{sec:previous}

Here we briefly compare with previous work on the same topic. The first closely
related paper is by Righi et al. (2008). Compared to this paper, the main differences with the present paper are
that we: consider the three-dimensional CO power spectrum rather than the angular power spectrum as a 
function of spectral resolution, we calculate the 21 cm-CO cross power spectrum, and that we consider
the detectability of both signals. In terms of modeling, Righi et al. (2008) adopt a more detailed model
for star formation that explicitly ties star formation to halo mergers. A second difference, mentioned already in 
\S \ref{sec:lum_co}, is that they calibrate their assumed $L_{\rm CO}-SFR$ relation to M82, while we normalize
at significantly lower SFRs. At $z=7$, our spatially averaged brightness temperature in the CO(2-1) agrees with
their results to within a factor of a few. Our larger brightness temperature is mainly driven by normalizing
the $L_{\rm CO}-SFR$ relation at
lower SFR. The two works are in general agreement given the model uncertainties.

The next related work is Carilli (2011). This work estimated the mean CO brightness temperature by using the
critical SFRD, along with empirical correlations between $L_{\rm CO}$, $L_{\rm FIR}$ and $SFR$. Indeed, we
followed Carilli (2011)'s lead by using the critical SFRD to constrain our modeling. Carilli (2011) finds brightness
temperatures of around $1 \mu K$ in the CO(1-0) and CO(2-1) lines (assuming the optically thick, high temperature
limit). This is in good agreement with our own estimates.\footnote{In detail, we make slightly different
assumptions in calculating the mean brightness temperature and so the near perfect agreement with our fiducial model is accidental. Specifically, we conservatively assume the same $L_{\rm CO}-SFR$ relation for the CO(2-1) and CO(1-0) lines, but normalize using the sub-linear scaling of $L_{\rm CO}$ with $SFR$. Carilli (2011), on the other hand, assumes a linear $L_{\rm CO}-SFR$ relation throughout but uses the optically thick, high temperature limit to infer the brightness
temperature of each line. Nonetheless, the two calculations are in broad agreement given the uncertainties.} We expand on Carilli's (2011)
work by calculating the spatial fluctuations in the CO emission, the cross power spectrum with the redshifted
21 cm signal, and the detectability of each signal.

Finally, the work with the closest overlap to this paper is that of Gong et al. (2011). These authors find
a spatially averaged brightness temperature, in the CO(1-0) line, of order $1 \mu K$ at $z \sim 6-7$. Taken
at face value, this is broadly consistent with our findings. Interestingly, however, their assumed $L_{\rm CO}(M)$
relation is rather different than our fiducial model and the models of Righi et al. (2008); their model weights the high mass halos much more strongly (see their Figure 1). Additionally, it appears that $f_{\rm duty} \sim 1$ in their model in contrast with our fiducial model. 
We have verified that we reproduce their mean brightness temperature if we adopt the same $L_{\rm CO}-M$ relation.
Their $L_{\rm CO}-M$ relation is extracted from the Obreschkow et al. (2009) model for CO luminous galaxies. We discussed our reservations about using this model directly in \S \ref{sec:lum_co}. It is not clear what the main
source of the difference between their $L_{\rm CO}-M$ relation and our models are: presumably the Obreschkow et al. (2009) model matches the empirical $L_{\rm CO}-SFR$ relations at intermediate redshift that are used to calibrate
our models. The difference then must lie in extrapolating this to higher redshift, where semi-analytic models
have yet to be tested, and to small halo masses that are unresolved in the Obreschkow et al. (2009) calculations.
It would be interesting to investigate this in more detail. 

\section{Conclusions}
\label{sec:conclusions}

We have quantified the possibility of intensity mapping observations in rotational emission lines from CO molecules during the EoR. We model both the power spectrum of
CO emission fluctuations, and the cross-correlation between CO and redshifted 21 cm signals. These measurements would be extremely interesting for three reasons.
First, the CO measurement would directly inform models for the properties of molecular clouds -- the sites of star formation -- in the very galaxies that reionize the Universe. Second,
the cross spectrum would provide a more direct measure of the size of ionized regions around the ionizing sources than available from the redshifted 21 cm signal alone.
Third, the cross spectrum would help confirm the high redshift nature of a putative redshifted 21 cm signal.

We find that the signal is potentially detectable with a relatively modest-sounding interferometer for the CO survey. Quantitatively, at $z=7$ our fiducial model predicts a spatially-averaged brightness temperature 
of $\avg{T_{\rm CO}} \sim 1 \mu K$, brightness temperature fluctuations of amplitude $\Delta^2_{\rm CO, CO} = 0.2 (\mu K)^2$
at $k=0.1 h$ Mpc$^{-1}$, and a 21 cm-CO cross power spectrum at the level of $\Delta^2_{\rm 21, CO} (k) \sim 10^3 (\mu K)^2$
at $k=0.1 h$ Mpc$^{-1}$ if the Universe is $\sim 80\%$ ionized at this redshift.
The 
theoretical forecasts are, however, uncertain because
we have only a limited empirical and theoretical handle on the properties of CO emitting gas in the low-luminosity, high redshift galaxies of interest. 
Here we mention a few possible approaches to improving this situation:
\begin{itemize}

\item Early data from ALMA will presumably constrain some of the uncertainties in our modeling.

\item Chang et al. (2011) have detected the cross-correlation between a post-reionization 21 cm intensity map and optically selected galaxies from the DEEP2 survey at $z \sim 1$. A powerful
proof of principle would be to carry out a similar cross-correlation between a low-redshift CO intensity map, low redshift 21 cm emission, and optically selected galaxies. In addition to being a strong demonstration
of the feasibility of these measurements, this measurement would provide constraints on the aggregate properties of CO emission at low redshift where star formation is more well understood. It would also
serve to address any worries about correlated foregrounds in the 21 cm and CO maps. A similar possibility is to cross correlate low redshift CO maps with FIRB maps from Planck
and Herschel (e.g. Ade et al. 2011); this would also help constrain which source redshifts dominate the FIRB emission.

\item Yet another possible cross-correlation is to combine high redshift CO maps with LAE surveys. Since the properties of individual LAEs are fairly well determined, this will help constrain the nature of the CO emitting galaxies in a relatively model independent way. 

\item Further theoretical work is needed to help understand plausible values of $T_{\rm ex}$ at high redshift, and its dependence on gas density, the CMB
temperature, and on the sources of heat input at high redshift.

\end{itemize}

It would also be interesting to consider intensity mapping with other emission lines (e.g. Suginohara et al.
1999, Hern\'andez-Monteagudo et al. 2007, 2008, Visbal \& Loeb 2010). The CII 158 $\mu$m line is the most
luminous and perhaps most promising other line to consider; the advantage of CO is that one can potentially
measure two rotational lines, providing an important systematics check, and that the necessary millimeter wavelength
observations for high redshift CII lines are more challenging than observations at centimeter-scale wavelengths.

In conclusion, we believe that CO intensity mapping is a potentially powerful approach to measuring large scale structure at high redshift. Further theoretical and observational work is
needed to address the uncertainties in predicting the CO signal, but we have outlined several possible approaches to improve this situations. Ultimately, when used in conjunction with
measurements of the redshifted 21 cm signal, CO intensity mapping may provide a powerful probe of the EoR and high redshift star formation.

\section*{Acknowledgments}
We thank Matt McQuinn for providing the reionization simulations used
in this analysis and for comments on a draft. This work was initiated at the summer 2010 Aspen 21 cm
cosmology meeting. This meeting and a subsequent meeting at the Keck
Institute for Space Sciences helped fuel this work, and we acknowledge
useful conversations with the participants of these meetings. We are especially grateful to Judd Bowman as a co-organizer of both of these
meetings.
SRF was partially supported by the David and Lucile Packard Foundation, by the Alfred P. Sloan Foundation, and by NASA through the
LUNAR program. The LUNAR consortium (http://lunar.colorado.edu), headquartered at the University of Colorado, is funded by the NASA
Lunar Science Institute (via Cooperative Agreement NNA09DB30A) to investigate concepts for astrophysical observatories on the Moon.
SPO acknowledges NSF grant AST 0908480 for support. Part of the research described in this paper was carried out at the Jet Propulsion Laboratory,  California Institute of Technology,
under a contract with the National Aeronautics and Space Administration.



\end{document}